\documentclass[aps,pre,twocolumn,superscriptaddress,floatfix,amsfonts,citeautoscript]{revtex4}
\usepackage{times}
\usepackage{graphicx}
\usepackage{graphics}
\usepackage{dcolumn}
\usepackage{bm}
\usepackage{color}
\usepackage{amssymb} 
\usepackage{amsmath}
\begin{document}

\title{\textbf{Static and Dynamic Properties of Two Dimensional Coulomb Clusters}}

\author{Biswarup Ash}
\affiliation{Indian Institute of Science Education and Research-Kolkata, Mohanpur, India-741246}

\author{J. Chakrabarti}
\affiliation{S.N. Bose National Centre for Basic Sciences, Block-JD, Sector-III, Salt Lake, Kolkata-700098}

\author{Amit Ghosal}
\affiliation{Indian Institute of Science Education and Research-Kolkata, Mohanpur, India-741246}

\begin{abstract}
We study the temperature dependence of static and dynamic responses of Coulomb interacting particles in two-dimensional traps across the thermal crossover from an amorphous solid- to liquid-like behaviors. While static correlations, that investigate the translational and bond orientational order in the confinements, show the footprints of hexatic-like phase at low temperature, dynamics of the particles slow down considerably in this state -- reminiscent of a supercooled liquid. Using density correlations, we probe intriguing signatures of long-lived inhomogeneities due to the interplay of the irregularity in the confinement and long-range Coulomb interactions. The relaxation at multiple time scales show stretched-exponential decay of spatial correlations in irregular traps. Temperature dependence of characteristic time scales, depicting the structural relaxation of the system, show striking similarities with those observed for the glassy systems indicating that, some of the key signatures of supercooled liquids emerge in confinements with lower spatial symmetries.
\end{abstract}

\maketitle

\section{INTRODUCTION} 

Long-range interacting classical particles in disordered media can give rise to exotic phases, e.g, hexatic glass~\cite{Chudnovsky}, which is characterized by short range positional order and quasi-long range orientational order. This is quite similar to the usual hexatic phase in two dimension ($\mathrm{2D}$), as discussed in Kosterlitz, Thouless~\cite{KT1,KT2}, Halperin, Nelson, and Young (KTHNY)~\cite{HN1,HN2,Young} melting scenario. Though a true hexatic glass in the thermodynamic limit attracted critical debates~\cite{Toner1991,Toner3}, it may be realized in finite systems~\cite{Toner2}. However, there have been attempts made to understand experiments with colloidal particles~\cite{Kusner1994}, binary-alloy~\cite{Zippelius}, magnetic bubble lattice~\cite{Seshadri} and disordered type II superconductors~\cite{Chudnovsky} in light of a hexatic glass.

Confined systems, with small number of particles, are of great current interests due to their controlled experimental tunability over a wide range of parameters. Such systems are not only significant from technological perspective, they are also very important for fundamental Physics: They are ideal play ground for exploring complex interplay of interactions and disorder. In this paper, we develop an understanding of such interplay by studying static and dynamic correlations in nano-clusters with (long-range) Coulomb-interacting particles.

Standard order-disorder transitions do not occur in $\mathrm{2D}$~\cite{MerminWagner} because, a true ordered phase is weakened by fluctuations producing only quasi-ordered state. Still a transition from solid to liquid occurs with an intervening hexatic phase which is characterized by short range positional order and quasi-long range bond-orientational order. Such two step melting scenario, predicted by the KTHNY theory for clean $\mathrm{2D}$ systems, is realized in certain experiments~\cite{Y_Han,G_Maret2004,Peng} and simulations~\cite{Prestipino,Krauth2015}. The possibility of extending such ideas in the presence of disorder is still a subject of current research~\cite{Nelson_disorder, Serota, Peter2013, Lyuksyutov}. Does the melting of Coulomb interacting particles in $\mathrm{2D}$ system corroborate with the KTHNY melting, even in the absence of any disorder? In spite of thirty years of research, there is no definitive answer, to the best of our knowledge~\cite{Mazars2015}. 

Study of dynamics across the melting often provides crucial additional insights~\cite{Zahn_Maret_PRL2000}. The dynamical behavior of a system indeed helps in probing the intricacies of different thermal phases. The dynamical response of $\mathrm{2D}$ systems, close to the liquid-hexatic transition, show some striking similarities with that of liquids close to the glass transition~\cite{Ediger_2000}. Molecular dynamics simulation of $\mathrm{2D}$ system of colloid particles~\cite{Zangi} and experiment with granular materials~\cite{Dauchot,Reis_2007} show that with the increase in orientational order, the dynamics of the particles slow down and become heterogeneous. Particles undergo `caging effect' (i.e., getting  temporarily  trapped in the cage of its neighbors) and cooperative dynamics. These results into sub-diffusive behavior of the mean square displacements as observed in experiments~\cite{Weeks627} and simulations~\cite{Kob_Andersen1,Kob_Andersen2,Kob_PRL_String}.

While true phase transitions pertain to bulk systems only, for finite systems the notion of solid- and liquid-like `phases' have been used successfully to characterize qualitative behaviors of macroscopic systems~\cite{Bonitz_2008}. In recent years, static and dynamic properties of $\mathrm{2D}$ finite clusters were studied for different types of confinements~\cite{BP94,DA13,DA16, EPL2016, Peeters_NJP} and interaction potentials~\cite{Ptr_anistrpy_log, DA13, Yukawa_Theory} across the thermal melting. These theoretical studies are motivated by experimental realizations, such as colloidal suspensions~\cite{2D_colloid_exp, Colliod_simltn, Mag_part}, confined plasma~\cite{Melzer_plasma_Exp, Melzer_plasma_Simulation, plasma_exp}, electrons in quantum dots in high magnetic fields~\cite{QD_Ashoori,QD}, radio-frequency ion traps~\cite{RF_trap} and electrons on the surface of liquid helium~\cite{He_wigner}. Recent studies of static~\cite{DA13} and dynamic~\cite{EPL2016} properties of Coulomb interacting particles in irregular geometry have confirmed the thermal crossover from `solid' to `liquid'-like phases. From the study of spatio-temporal correlations~\cite{EPL2016} in such system, it is found that with decrease in $T$ some of the key signatures of glassy dynamics emerges. For example, as system approaches solidity from liquid-like phase, the dynamics of the particles become slow and heterogeneous. Is this a signature of a hexatic-glass phase?

We address this broad question by presenting in this paper the results of extensive computer simulations on the static and dynamic properties of Coulomb interacting particles in different confinement geometries. From our analysis of static properties, we find that while the positional order is depleted even at the lowest $T$ for an irregular trap, a solid-like phase can still be identified at low $T$ due to the bond orientational order. A liquid-like phase emerges~\cite{DA13} with the breaking down of bond orientational order beyond a crossover temperature $T_X$, however, the inhomogeneities in the liquid persists up to a much larger temperature -- about an order of magnitude larger than $T_X$. Beyond this large temperature scale ($\sim 10T_X$), the system crosses over to a standard isotropic liquid. Our results are indicative of a hexatic-like low $T$ phase in the confinements. Finally, the analysis of the trajectories of individual particles and the $T$-dependence of relaxation time-scales show characteristics quite similar to those observed for glass formers near the glass transition temperature~\cite{Ediger_2000}.

The rest of the paper is organized as follows: In Sec. II we will give the details of the models and methods used in our simulations emphasizing the way disorder is introduced in finite systems. We will also discuss about the methods used to study the system. Then we present, in Sec. III, our results for static properties where we analyze the thermal crossover through the temperature dependence of bond orientational order. In Sec. IV dynamic properties are analyzed to identify, in particular, any signature of a glass-like behavior, and also to study the temperature dependence of structural relaxation time(s). Finally, we conclude in Sec. V.

\section{MODEL and Method}

We consider $N$ classical particles each with charge $q$ in a confining potential $V_{\rm conf}(r)$. These particles interact via long range Coulomb potential and are restricted to move in $\mathrm{2D}$. The Hamiltonian describing such a system reads as:
\begin{equation}
{\cal H} = \frac{q^2}{\epsilon} \sum_{i<j=1}^{N} \frac{1}{|\vec{r}_i - \vec{r}_j |} + \sum_{i=1}^{N} V_{\rm conf}(r_i),
\end{equation}
 where, $r_i = \vert \vec{r}_i \vert = \sqrt{x_i^2 + y_i^2}$ is the distance of the $i$-th particle from the origin. Here, first term in the Hamiltonian represents Coulomb repulsion between particles in a medium with dielectric constant $\epsilon$. We consider long range Coulomb repulsion alone, because systems with small number of particles, particularly, in the presence of disorder, are expected to offer a very weak screening, if at all. In our study we considered two types of confinement potential; (a) parabolic (having circular symmetry):
\begin{equation}
 V_{\rm conf}^{\rm Cr}(r)= \alpha r^2, ~~~\text{ where}~~~ \alpha = m \omega_0^2/2,
\end{equation}
 and (b) Irregular (that lacks all spatial symmetries)~\cite{Bohigas93, DA13}:
\begin{equation}
V_{\rm conf}^{\rm Ir}(r)=a\{ x^4/b+by^4-2\lambda {x^2}{y^2} +\gamma (x-y)xyr\}. 
\end{equation}
The solid phase of Coulomb clusters at low $T$ is called a Wigner Molecule (WM)~\cite{R_Egger}, because it mimics the physics of Wigner crystal~\cite{Wigner1934}. In finite systems, we refer a WM as circular Wigner molecule (CWM) in a circular trap, and as irregular Wigner molecule (IWM) in an irregular trap.

We rescale the length $r' \rightarrow \phi^{1/3}\alpha^{-1/3} r$ and energy $E' \rightarrow \phi^{2/3}\alpha^{1/3} E$, where $\phi = q^2/\epsilon$,  in such a way that the CWM Hamiltonian transforms to~\cite{BP94}:
\begin{equation}
 {\cal H}^{\rm Cr} = \sum_{i<j=1}^{N} \frac{1}{|\vec{r}_i - \vec{r}_j |} + \sum_i r_i^2 
\end{equation}
correspondingly the time scale will also be renormalized to $t'= \hbar~\phi^{-2/3}\alpha^{-1/3} t$. Such renormalization of length, energy and time will eventually make temperature to be expressed as $T'=E'/k_B$, where $k_B$ is the Boltzmann constant. In order to have an estimate of these new length and time scales in conventional units, we consider electrons in GaAs heterostructure with a typical confinement energy of $\hbar \omega_0 = 1$ meV. The above scaled length $(r')$, energy $(E')$ and time $(t')$ unit takes the value of $630$ \AA, $1.7$ meV and $376$ fs, respectively. In extracting these values we have used, $q = -e$, the charge of an electron, $m = 0.067 m_e$, the mass of an electron and $\epsilon = 13 $~\cite{parameters}. 

$ V_{\rm conf}^{\rm Cr}$ is quadratic in length-scale while $V_{\rm conf}^{\rm Ir}$ is quartic. In order to facilitate a justified comparison between the two, we need to express the quartic irregular confinement in the units of quadratic circular confinement. This is achieved by setting the parameter $a =\left(m\omega_0^2/2r'^2\right) a'$, which includes both $\omega_0$ and a scaling factor $a'$, for irregular confinement. Thus a judicious choice of $a'$, which now controls the strength of the irregular confinement, brings the two confinements on equal footing on dimensional ground. Following the same scale transformations as for the circular case, we obtain the following Hamiltonian for $V_{\rm conf}^{\rm Ir}$:

\begin{eqnarray}
{\cal H}^{\rm Ir} &=& a'\left[\frac{x^4}{b}+by^4-2\lambda {x^2}{y^2} +\gamma (x-y)xyr\right] \nonumber \\
  &+& \sum_{1 \leq i < j}^{N} \frac{1}{|\vec{r}_i - \vec{r}_j|}
\end{eqnarray}

There is still a parameter, $a'$, left to be fixed for $V_{\rm conf}^{\rm Ir}(r)$. For a given $N$, we choose the value of  $a'$ according to the following prescription: We know that the model of a $\mathrm{2D}$ electron gas, neutralized by a uniform positive background, and interacting via Coulomb interaction can be characterized by a single dimensionless quantity $\Gamma = \sqrt{\pi n}/T$, where $n$ is the average number density~\cite{Gan_Chester}. Thus, for a given $T$, the thermodynamic properties of the system is completely determined by its density, $n$. In the present context of finite systems, we adopt the same view and assume that our systems are also characterized by the same parameter $\Gamma$ (which we will justify below). In order to make a meaningful comparison between systems in different confinements but at same $T$ and with same $N$, we fix the density by tuning $a'$ for irregular confinement and $\alpha$ for circular confinement. The assumption of single parameter description in terms of $\Gamma$ is later indicated (in sec. IIIC) by its bulk value at melting, irrespective of $N$ and trap geometries. Note that $a'$ tunes the average density by making the quartic oscillator narrow or shallow. For a given $N$, we fixed the value of $a'$ for the irregular confinement in such a way that the average inter-particle distance is equal to that for circular confinement at the lowest $T$. In this paper, we report results for $N=75, 150$ and $500$ number of particles which require $a'=0.10,0.055$ and $0.020$, respectively, following above procedure. 

Note that while the thermal crossover in circular trap is already studied in great details~\cite{BP94}, the main objective of this paper lies with the understanding of melting in irregular confinement. We study circular confinement mainly for comparison and thus move those results predominantly in supplementary materials. 

The parameter $\lambda$ in irregular confinement controls chaoticity in the single-particle dynamics~\cite{Bohigas93}. Tuning $\lambda$ from zero to unity generates periodic to chaotic motion of a sole particle in the trap. Chaotic motion, along with broken spatial symmetries are taken as the signatures of disorder in our study. The parameter $b (=\pi/4)$ breaks the symmetry of a square and $\gamma$ breaks the reflection symmetry. We consider $\lambda \in [0.565,0.635]$ and $\gamma \in [0.10,0.20]$~\cite{Hong03}. The values of different parameters are adjusted to generate self-similar copies of the system over which statistics are collected for the purpose of ``disorder averaging" of physical observables~\cite{Ullmo03}. Results from CWM are averaged for many independent simulations. 

To study the static properties, we carried out (classical) Metropolis Monte Carlo (MC) simulation~\cite{FrenkelBook} aided by simulated annealing algorithm~\cite{Kirkpatrick}.  Dynamical responses are studied using molecular dynamics simulation (MD)~\cite{FrenkelBook}. To achieve a desired $T$, we have used Berendsen like thermostat 
~\cite{FrenkelBook} during the equilibration. Once equilibration is done, we have used conventional velocity-Verlet algorithm~\cite{FrenkelBook} to integrate the equations of motion. We have performed MD runs up to $2\times 10^6$ steps with a time step size of $dt = 0.005t'$. We find excellent match of different observables obtained from Monte Carlo simulation with the time-averaged results of those quantities obtained from molecular dynamics simulation, except for the lowest temperatures featuring slow dynamics, as we discuss below. Such match of the physical results from the two independent methods of numerical calculations validate the correctness of our findings.

\section{STATIC PROPERTIES}

The broken symmetry state of a system at low $T$ can be identified by its order. For example, crystalline solid is characterized by the long range positional and orientational order. We thus proceed to explore first the positional order in our confined systems at the lowest $T$. Note that even though translational symmetry is broken by confinements, a circularly confined system would still possess an azimuthal periodicity~\cite{BP94}. 
 
\subsection{POSITIONAL ORDER}
\begin{figure}[t]
\includegraphics[width=8.5cm,keepaspectratio]{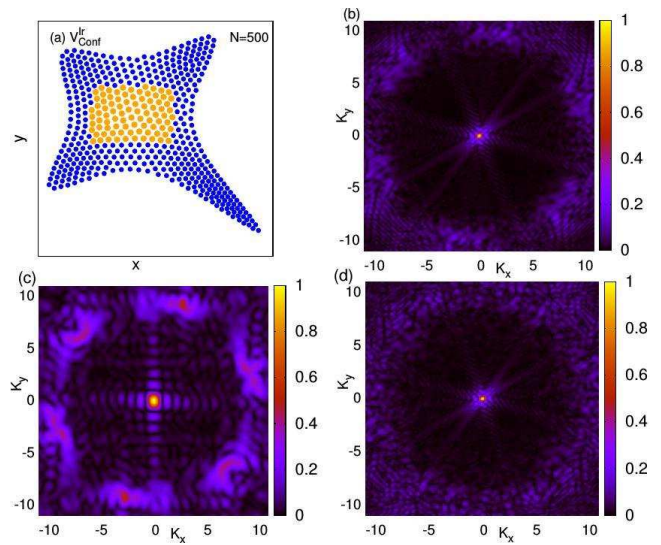}
\caption{
	  (a) The ground state $(T=0)$ configuration of $N=500$ particles in irregular trap ($\lambda=0.635, \gamma=0.20$). The orange dots are selected central particles for the search of positional order (See Sec. IIIA), while all other particles are represented by blue dots. Panel (b)-(d) show Fourier transform of particle density for various cases: Panel (b) is for the ground state with all particles shown in panel (a). Panel (c) is for the orange particles only in the central region of panel (a). Panel (d) is for all particles, but at $T=0.02 \sim T_X$. In panel (b-d) magnitude of $\rho(\vec{k})$ is scaled to unity for visual clarity.
}
\label{fig:IWM_FT}
\end{figure}

In order to quantify the positional order in IWM, we plot the Fourier component of local density, $\rho(\vec{k}) = \sum_{j=1}^{N} \mathrm{exp}[i \vec{k} \cdot \vec{r}_j]$, in Fig.~\ref{fig:IWM_FT}(b). Here, $\vec{k}$ is the momentum vector and $\vec{r}_j$ represents the position vector of the $j$-th particle. In all analysis, distances are measured in the unit of $r_0$, the mean inter-particle spacing, at the lowest $T$, between neighboring particles (which is same by construction for irregular and circular confinements, for a given $N$). 
For a perfect crystalline structure, $\rho(\vec{k})$ consists of Bragg peaks at the reciprocal lattice vectors and such peaks broaden with $T$, finally disappearing upon melting. 
Fig.~\ref{fig:IWM_FT}(a) shows the ground state configuration, obtained following the prescription of Ref.\cite{DA13}, of $N=500$ particles in irregular confinement (with $\lambda=0.635, \gamma=0.20$) and Fig.~\ref{fig:IWM_FT}(b) shows $\rho(\vec{k})$ corresponding to this configuration.
We find no strong peak in reciprocal space, apart from the one at $\vec{k}=0$. Within the diffuse pattern at $\vec{k}\ne0$, we broadly identify six patches. 
Broadness of these patches ensures that there is, at best, a very short range (much shorter than the system size) positional order, if any, in IWM. We can probe this order further by considering the fact that the ground state configuration for a trapped system is a result of the interplay between the formation of a triangular lattice (minimum energy configuration for a bulk $\mathrm{2D}$ Coulomb system~\cite{Bonsall}) towards the central part of the trap (i.e., far away from the boundary) and the geometry of the confining potential near the boundary.
In search of any positional order deep inside the system, we consider only those particles which belong to a small region (of length $l_x=7 r_0$ along $x$-axis and $\sqrt{3}l_x/2$ along $y$-axis, to accommodate a commensurate triangular lattice structure) near the center of the confinement. These particles are colored orange in Fig.~\ref{fig:IWM_FT}(a). We then calculate $\rho(\vec{k})$ considering only the orange particles. Fig.~\ref{fig:IWM_FT}(c) shows that the six patches become somewhat prominent enhancing positional order in this subsystem. This is also obvious from Fig.~\ref{fig:IWM_FT}(a), where the lattice lines become wiggly even in the subsystem due to irregular geometry. At $T =0.020$, which characterizes the crossover temperature (as discussed later), $\rho(\vec{k})$ shows a diffuse pattern ( Fig.~\ref{fig:IWM_FT}(d) ), similar to that at $T=0$, establishing little evolution of positional order, if any, with temperature. Similar analysis for the CWM shows stronger positional order, primarily reflecting the azimuthal symmetry, though it gets depleted upon averaging over whole system (see Fig.~1 in supplementary materials). 

In the absence of any appreciable positional order, any solidity in IWM is contributed by the orientational order in 2D. Motivated by this assertion, we next analyze the bond orientational order in the system.

\subsection{BOND ORIENTATIONAL ORDER (BOO)}

In Fig.~\ref{fig:IWM_FT}(a) we see that particles which are not on the boundary are mostly surrounded by six nearest neighbors oriented in a hexagonal fashion and thus indicate six-fold bond orientational order (BOO) of a triangular lattice. In $\mathrm{2D}$ systems, the BOO parameter, $\psi_{6}(k)$, for $k$-th particle is defined as~\cite{NelsonBook02}:
\begin{equation}
\psi_{6}(k) = \frac{1}{N_b} \sum_{l=1}^{N_b} \mathrm{e}^{i6\theta_{kl}}
\label{Eq:BOO}
\end{equation}
Here, $\theta_{kl}$ is the angle made by the bond between the particles $k$ and $l$ with an arbitrary axis and $N_b$ is the number of nearest neighbors of particle $k$. We identify the nearest neighbours of a given particle by Voronoi construction~\cite{Voronoi}. $\left\vert \psi_{6}\right\vert$ achieves a maximum value of unity in a perfect triangular lattice and would get smaller with distortions. 

\begin{figure}[t]
\includegraphics[width=8.5cm,keepaspectratio]{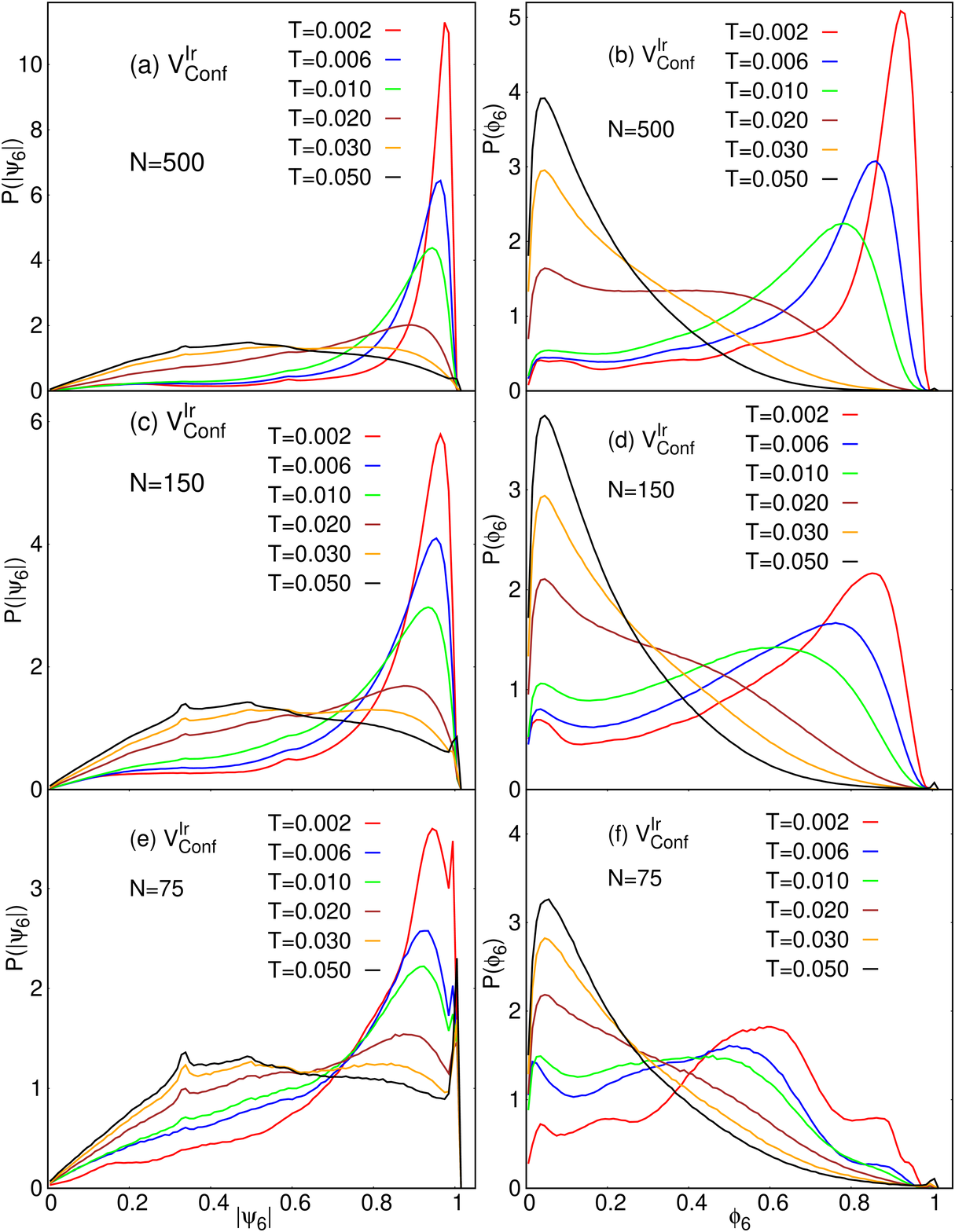}
\caption{
	 Evolution of the bond orientational order with $T$ in irregular confinement is shown for $N=500$ in panel (a), for $N=150$ in panel (c) and for $N=75$ in panel (e). The low-$T$ peak in $P(|\Psi_6|)$ at $|\Psi_6| \sim 1$, that signifies the ``solidity", smears out as the system undergoes the thermal crossover. The small and weak peak in $P(|\Psi_6|)$ at $|\Psi_6| \sim 1$ at high $T$, particularly for smaller $N$, is a spurious one (see text) and does not constitute BOO.
Similar distribution of $\phi_6$ representing the bond orientational correlation up to nearest neighboring distance (see text) are shown in panels (b, d, f). $P(\phi_6)$ features a bimodal structures, its peak near $\phi_6 \sim 1$ (at low $T$) signaling solidity and the one at $\phi_6 \approx 0$ portraying liquidity.
}
\label{fig:psi61}
\end{figure}

Fig.~\ref{fig:psi61}, shows the $T$-dependence of the probability distribution, $P(\left\vert \psi_{6} \right\vert)$, of $\left\vert \psi_{6} \right\vert (= \left\vert\psi_{6}(k)\right\vert$ for $k=1,2, \cdots N)$ for IWM with $N=500$ (panel a), $150$ (panel c) and $75$ (panel e) particles. For $N=500$ (Fig.~\ref{fig:psi61}(a)), this distribution is sharply peaked around unity at low $T (=0.002)$, describing an orientationally ordered phase at low $T$. With increase in $T$, the peak broadens due to thermal fluctuations that distorts the bond angles $(\theta_{kl})$ with the neighbors. For $T>0.02$, we get a very broad distribution signifying the breakdown of orientational order. Thus, the $T$-dependence of $P(\left\vert \psi_{6}\right\vert)$ signals a thermal crossover from an orientationally ordered solid-like to a disordered liquid-like phase in confined systems. It is interesting to note that $P(\left\vert \psi_{6}\right\vert)$ shows hardly any thermal evolution for $T > 0.03$. This is found true irrespective of the confinement geometry or total number of particles,$N$. This insensitivity confirms that the scrambling of orienational order and hence melting is complete by $T\approx 0.03$. 

With decreasing $N$ (Fig.~\ref{fig:psi61}(c) for $N=150$ and Fig.~\ref{fig:psi61}(e) for $N=75$), we notice a strengthening of a small and feeble peak in $P(\left\vert \psi_{6} \right\vert)$ near $\left\vert \psi_{6} \right\vert=1$. We assert that such peak is spurious. We verified that this is contributed predominantly by the particles on the boundary, having very few nearest neighbors. Note that a particle with a sole neighbor must yield $\left\vert \psi_{6} \right\vert=1$, irrespective of the value of bond angle. Decrease in $N$ increases the ratio of the number of boundary- to bulk-particles, enhancing the strength of such spurious peak.
From our overall study, we find that systems with $N \le 100$ tend to show non-universal behavior, not only for $P(\left\vert \psi_{6} \right\vert)$ but also for other observables we describe below. On the other hand, systems with $N=100$-$500$ describe generic behavior featuring qualitatively similar results. 

The nature of melting by disrupting bond orientational order is usually tracked by studying the large $r$ behavior of bond-orientational correlation function, $g_6(r) =\langle \psi_{6}(r)\psi_{6}^{*}(0) \rangle$. However, for finite systems with small number of particles, $g_6(r)$ is not very useful, because any trend in its $r$-dependence is hard to discern~\cite{DA13} due to limited values of $r$.

\subsection{PROJECTION OF BOND ORIENTATIONAL ORDER}

In order to get  additional insight into the degree of local orientational order, we, next, look into the short distance projection of bond-orientational correlation function. The magnitude of the projection of bond orientational order, $\phi_6(k)$, of particle $k$ onto the average local bond orientational order of its nearest neighbours is defined as~\cite{Larson, Dillmann_2012}:

\begin{equation}
\phi_{6}(k) = \left\vert\psi_{6}(k)^{*}~ \frac{1}{N_b}\sum_{l=1}^{N_b} \psi_{6}(l)\right\vert
\end{equation}

While $\left\vert \psi_{6}(k) \right\vert$ measures how the neighbors of particle $k$ adjust locally on a hexagonal environment, $\phi_{6}(k)$ determines how the orientation of a particle fits into the mean orientational field of its nearest neighbors. Thus, $\phi_{6} (= \phi_{6}(k)$ for $k=1,2, \cdots N)$ represents the bond orientational correlation up to next nearest neighbor distances.

At low $T$, as most of the particles are surrounded by six regular neighbors, and we expect $\phi_6 \sim 1$. With increasing $T$, due to depleted orientational order, $\phi_6$ decreases and eventually becomes close to zero at high $T$. Thus, $\phi_6$ distinguishes a high $T$ phase clearly from a low $T$ phase.

\begin{figure}[t]
\includegraphics[width=8.5cm,keepaspectratio]{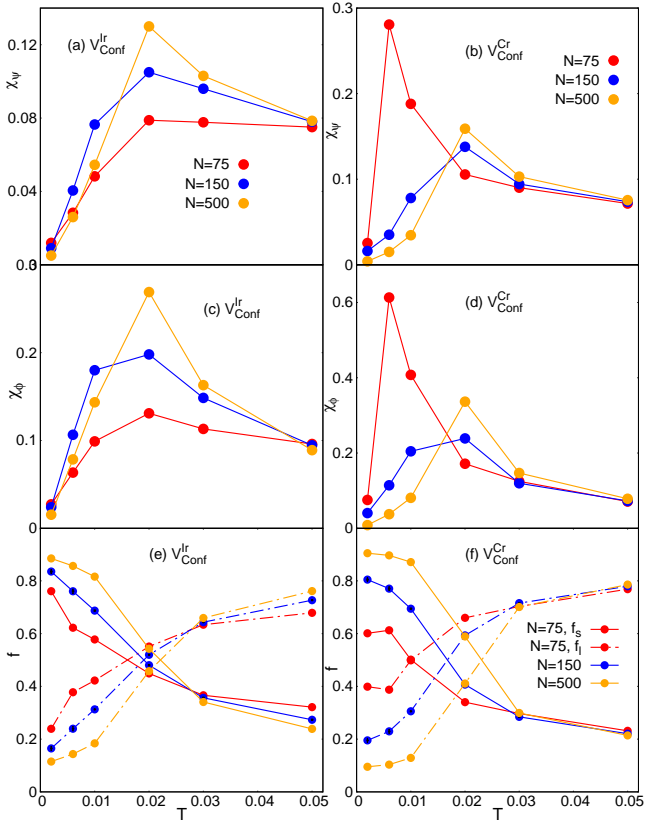}
\caption{
	  $T$-dependence of the generalized susceptibilities identifying the crossover temperature $T_X$, for $N=500, 150, 75$. Panel (a) and (c) shows the thermal evolution of $\chi_{\psi}$ and $\chi_{\phi}$ respectively in irregular confinement. While the peaks of $\chi_{\psi}$ and $\chi_{\phi}$ sharpen with $N$, their location, $T_X$, suffer hardly any change. Panel (b) and (d) shows results similar to those in panel (a) and (c) but for circular trap. The fragile order in this case for $N=75$, degrades $T_X$. Panel (e) and (f) illustrate the thermal evolution of the solid- and liquid-fraction in the system for circular and irregular trap. The crossing of the two takes place in a narrow window of $T$ (for different $N$) for $V^{\rm Ir}_{\rm conf}(r)$, where as, the temperature window is broad for $V^{\rm Cr}_{\rm conf}(r)$.
}
\label{Fig:opfluc}
\end{figure}

In Fig.~\ref{fig:psi61}(b,d and f) we show the distribution of $\phi_6$ at different $T$ for IWM with $N=500, 150$ and $75$, respectively. At low $T(=0.002)$, the distribution is sharply peaked at $\phi_6 \approx 1$ (see Fig.~\ref{fig:psi61}(b)), depicting a strong correlation of the orientational ordering of a particle with its nearest neighbours. As $T$ increases, such correlation reduces and at high $T(=0.050)$, the distribution becomes sharply peaked around $\phi_6 \sim 1$. Thus, for $T>0.02$, the local six-fold order is lost even at the second shell, representing a `liquid'-like state. We see that with decrease in $N$ (Fig.~\ref{fig:psi61}(d,f)), the sharpness of the peak around $\phi_6 \sim 1$ gets considerably reduced at the lowest $T(=0.002)$ as the effect of the boundary particles becomes more prominent for smaller system size. For $N=75$ (Fig.~\ref{fig:psi61}(f)), an orientational ordering is hard to discern up to the second neighbour even at low $T$.

An earlier investigation~\cite{DA13} of IWM with $N=150$ found $T_X\approx 0.02$ (for our parameters) by studying the $T$-dependence of specific heat and Lindemann ratio, among other quantities. Here we identify $T_X$ by studying the fluctuation of the orientational order parameters $\vert \psi_6 \vert$ and $\phi_6$, defined as
\begin{align}
\chi_{\psi} &= N \left[ \langle |\psi_6|^2 \rangle - \langle |\psi_6| \rangle^2 \right] \\
\chi_{\phi} &= N \left[ \langle  \phi_6^2  \rangle - \langle  \phi_6  \rangle^2 \right]
\end{align}
Here, $\chi$'s are the generalized susceptibility correspond to order parameter $\left\vert \psi_{6}\right\vert$ and $\phi_6$, respectively.
In Fig.\ref{Fig:opfluc}, we show the temperature dependence of $\chi_{\psi}$ (panel a,b) and $\chi_{\phi}$ (panel c,d) for different system sizes. While, for a bulk system, susceptibility diverges at the critical point, such divergence turns into a crossover in a finite system. The fluctuations in $\chi_{\psi}$ and $\chi_{\phi}$ attain their maxima at the crossover temperature $T_X$ distinguishing the low temperature solid- and high temperature liquid-like phases, as seen from Fig.~\ref{Fig:opfluc}(a-d). The peak at $T_X\sim 0.02$ becomes sharper with increasing $N$ (though for $N=75$, the circular confinement shows deviation), and we generally find the value of $T_X$ quite insensitive to the nature of confinement. Thus, $T_X$ estimated from bond orientational order parameter is in good agreement with its previous estimates~\cite{DA13}.

In order to compare $T_X$ with the corresponding melting temperature for bulk $\mathrm{2D}$ systems, we estimate the parameter $\Gamma (= \sqrt{\pi n}/T)$ at $T=T_X$. It is found that for bulk systems, melting occurs for $\Gamma \approx 137$~\cite{Gan_Chester}. In Table~\ref{Tab_Tx}, we show $\Gamma$ for different $N$ and found that $\Gamma$ remains close to $137$ within acceptable tolerance for the large $N$'s of our study. 

\begin{table}[h]
\centering
\renewcommand{\arraystretch}{1.5}
\begin{tabular*}{\columnwidth}{@{\extracolsep{\fill}}|l|l|l|l|l|l|l|@{}}\hline
\multicolumn{1}{|c|}{$N$} &
\multicolumn{3}{c|}{Irregular} &
\multicolumn{3}{c|}{Circular}\\ \hline
    &  $n$   &        $T_X$          & $\Gamma$ &  $n$  &       $T_X$   	& $\Gamma$  \\ \hline
75  & 2.42   &  0.017 $\pm$ 0.003    &   162    &  1.90 &  0.011 $\pm$ 0.001  	&  222      \\ \hline
150 & 2.53   &  0.019 $\pm$ 0.001    &   148	&  2.12 &  0.019 $\pm$ 0.002	&  136	    \\ \hline
500 & 3.01   &  0.021 $\pm$ 0.001    &   146	&  2.78 &  0.020 $\pm$ 0.001	&  148      \\ \hline
\end{tabular*}
\caption{Estimation of the parameter $\Gamma = \sqrt{\pi n}/T_X$, characterizes the melting in bulk $\mathrm{2D}$ systems, for irregular and circular confinements. For bulk systems $\Gamma \approx 137$ at the transition. Note that the average particle density $n$ is estimated by computing the area covered by $N$ particles in the respective confinements. }
\label{Tab_Tx}
\end{table}

Since $\phi_6$ is a projection of $\psi_6$, $\phi_6(k)\le \left\vert \psi_{6}(k) \right\vert$. Because maximum value of either of $\psi_6$ or $\phi_6$ is unity, $\phi_6(k) + \left\vert \psi_{6}(k)\right\vert \le 2$. Hence, we can estimate melting by describing each particle to be either `solid'- or `liquid'-like using the majority rule~\cite{Larson}: particle $k$ is solid-like if $\phi_6(k) + \left\vert \psi_{6}(k)\right\vert> 1$ and  liquid-like otherwise. In Fig.~\ref{Fig:opfluc}(e-f) we have shown the fraction of solid- and liquid-like particles at different $T$ for irregular and circular confinements, respectively. We see that at very low $T (=0.002)$ almost $90\%$ of the particles are solid like while at high $T (=0.050)$ particles are mostly liquid like. Interestingly, the two curves cross each other around the same $T_X$ as obtained from other quantities above. 

Few points deserve special mention here: A phase with depleted positional order but featuring an  orientational order in the scale of system size is indicative of a hexatic phase within the KTHNY description of bulk 2D melting. However, in systems with disorder, like the irregular confinement considered here, similar characteristics should identify a so called `hexatic glass' phase. This phase differs from regular hexatic phase in the sense that it supports both the thermal and quenched defects, whereas the hexatic phase in KTHNY theory allows only thermal defects. While the existence of a thermodynamic hexatic glass phase has been debated~\cite{Toner1991}, its signature in small finite systems, like IWM, can possibly be discerned as our results in preceding section suggest. 

\section{Dynamic properties}

A slow and heterogeneous motion of constituent particles is a hallmark of glassy systems. Below, we resort to dynamical characterization of different phases of our system in order to explore its glassy nature, if at all. Some aspects of intriguing slow and heterogeneous dynamics in IWM, primarily extracted from the analysis of Van Hove correlation function, are recently reported~\cite{EPL2016} by us. Here we expand on those results, firstly, by characterizing qualitatively the motional signatures in the traps, and subsequently, by extracting time scales associated with the structural relaxation causing the depletion of bond orientational order. 

\begin{figure}[t]
\includegraphics[width=8.5cm,keepaspectratio]{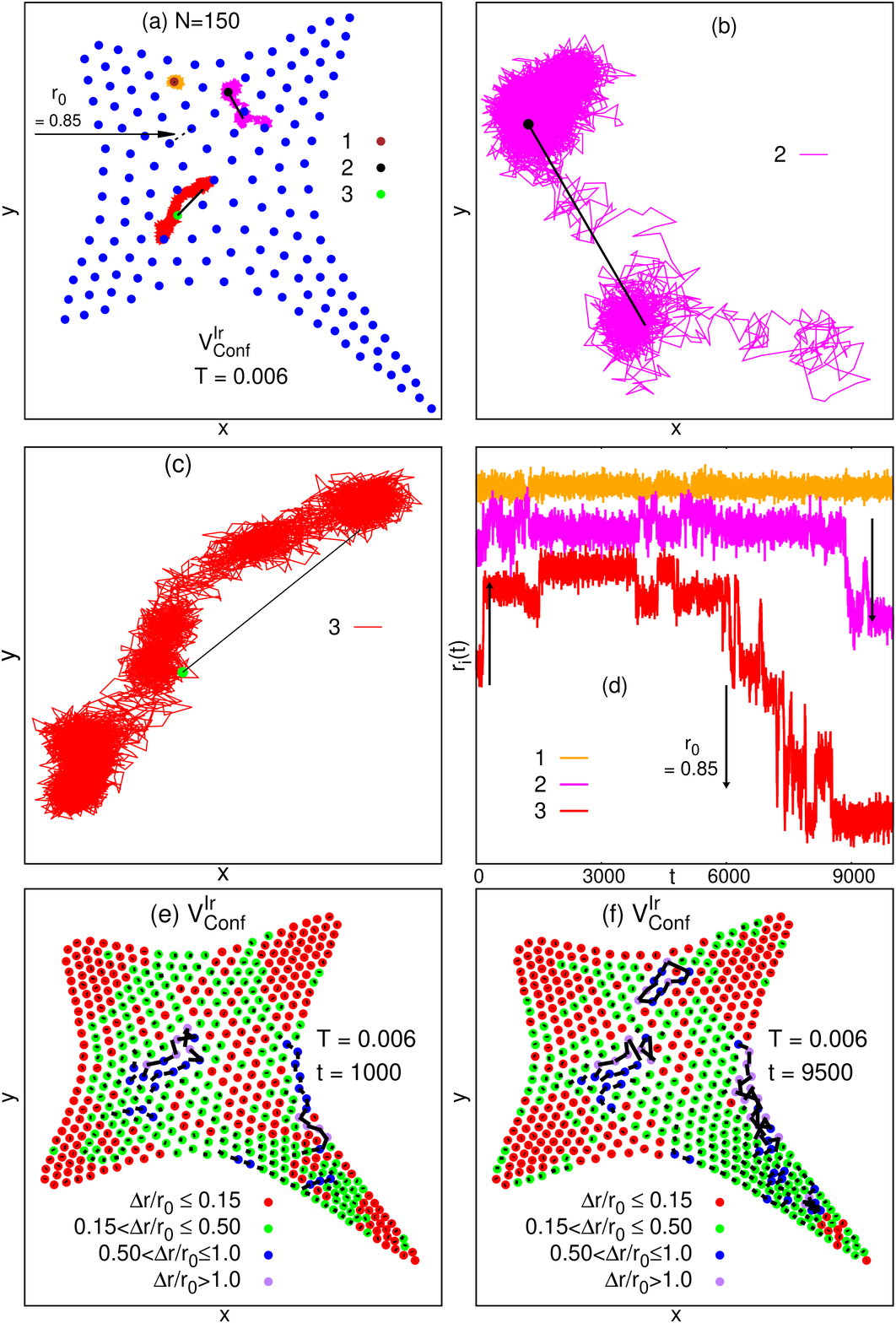}
\caption{
         (a) Initial positions of s$N=150$ particles (blue dots) in a specific realization of an  irregular confinement along with the trajectories of three particles at $T=0.006$. Thick black lines indicate the net displacement of the particles over the duration of the simulation. (b-c) Illustrates the trajectories of the particles $2$ and $3$ with larger displacements as shown in (a). Both of them jiggles around their equilibrium position for a while and they make occasional and sudden jumps by distance $\sim r_0$. Particle $1$, however, does not make such hops (not shown separately). (d) $t$-dependence of the instantaneous distance $r_i(t)$ from the centre of the confinement for the three selected particles $(i=1,2,3)$. Displacement ($\Delta \vec{r}_i(t) = \{\vec{r}_i(t) - \vec{r}_i(0)\}$) of $N=500$ particles in irregular confinement is shown for (e) $t=1000$ and (f) $ t=9500$ at $T=0.006$.
}
\label{Fig:IWM_Traj}
\end{figure}

\subsection{Analysis of the trajectory}

Focusing on a specific realization of the irregular confinement, $\{ \lambda, \gamma\}=\{0.635,0.20\}$, we show in Fig.~\ref{Fig:IWM_Traj}(a) the initial configuration of $N=150$ particles (thick dots) after MD equilibration. On this we also superimpose the trajectory of three particles (indexed as $1,2$ and $3$) during the whole MD-simulation. We see that, at low $T$, the nature of the dynamics of individual particles can be broadly classified in following two categories: \\
(i) There are particles, like the one indexed $1$ in Fig.~\ref{Fig:IWM_Traj}(a), which only rattles around their equilibrium position over the duration of whole MD simulation.\\
(ii) Additionally, there are particles, indexed $2$ and $3$ in Fig.~\ref{Fig:IWM_Traj}(a), which move longer distances (by several $r_0$) in the course of time. Their trajectories are enlarged in Fig.~\ref{Fig:IWM_Traj}(b) and (c) (along with a solid line representing the displacement). These particles jiggle around their equilibrium position for certain time and then jump close to location of its neighbor and spend some time before jumping to previous (reversible motion) or a new location (irreversible motion). Such occasional jumps can be captured by looking into the instantaneous position, $r_i(t)$ in Fig.~\ref{Fig:IWM_Traj}(d) where $i$ represent the particle index. We see that within a sufficiently small time window  these particles change their positions moving by a distance of the order of $r_0$. This motion can be regarded as the cage breaking process and avalanches of such cage-breaking events give rise to string-like pattern for the displacement $\Delta \vec{r}_i(t) = \{\vec{r}_i(t) - \vec{r}_i(0)\}$ of the particles as shown in Fig.~\ref{Fig:IWM_Traj}(e-f) for $N=500$. We, however, note that the length of such string in the scale of system size, decreases by going from $N=150$ to $N=500$. This cooperative motion in irregularly trapped Coulomb clusters is similar to those observed in other systems such as colloids~\cite{Nagamanasa_PNAS_2011}, Lennard-Jones mixture~\cite{Kob_PRL_String} and $\mathrm{2D}$ dusty plasma~\cite{Plasma_string}. Equivalent analysis for circular confinement (Fig.~3 in supplementary materials) also shows heterogeneous dynamics for the particles but the spatial heterogeneity in that case is dictated by its azimuthal symmetry. We conclude this section by asserting that similar motional footprints have been found for each realization (identified by a specific pair of $\{ \lambda, \gamma\}$) of irregular confinement we studied.

\begin{figure}[t]
\includegraphics[width=8.5cm,keepaspectratio]{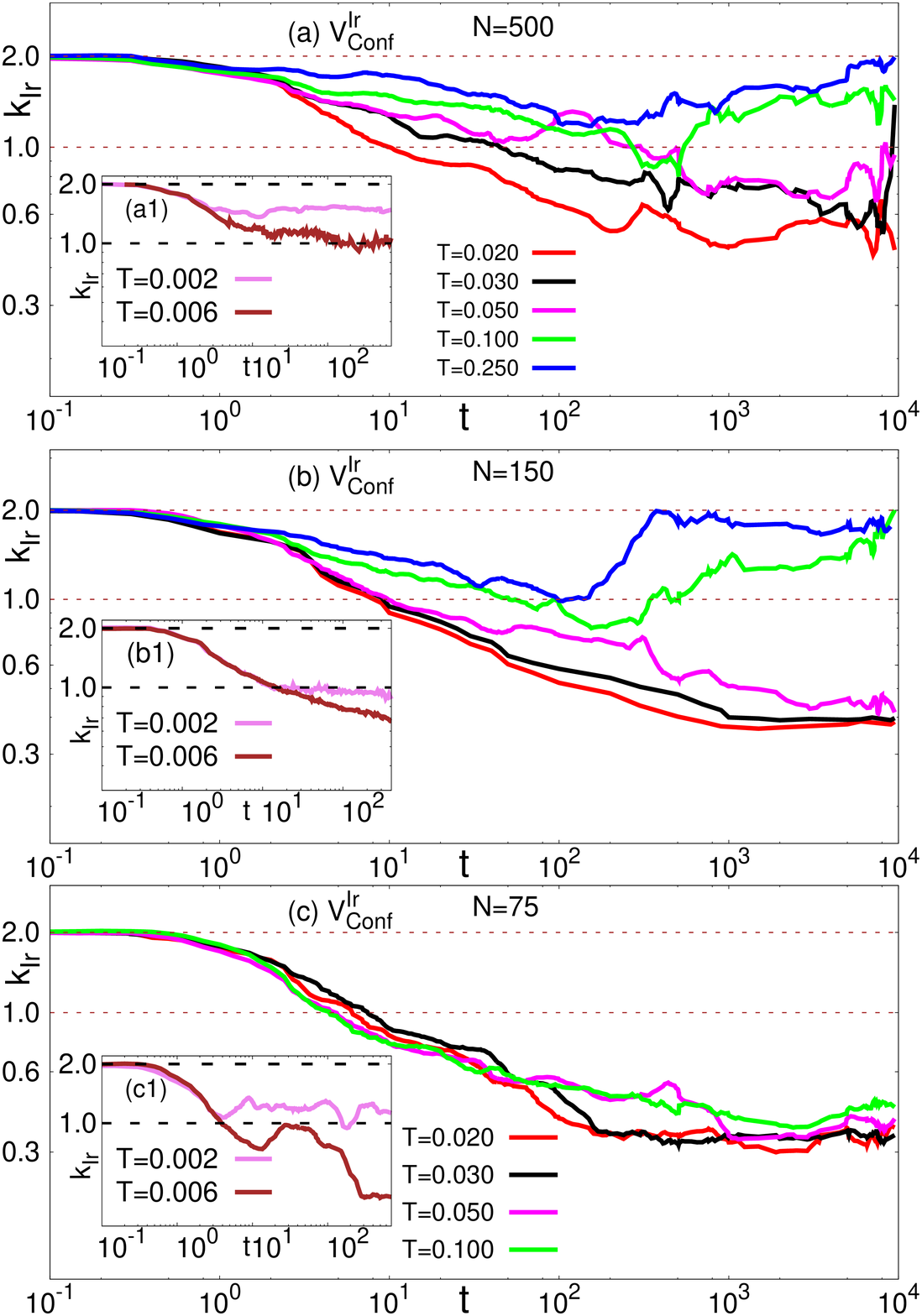}
\caption{
         Illustration of the intriguing spatial evolution of $G_s(r,t) \sim \mathrm{exp}[-l r^{k_{\rm Ir}}]$ by depicting the $t$-dependence of $\mathrm{k_{Ir}}$ (see text), at different $T$ for $V^{\rm Ir}_{\rm conf}$. The results are analyzed for (a) $N=500$, (b) $N=150$ and (c) $N=75$.
}
\label{Fig:SVH_Fit500}
\end{figure}

\subsection{Inhomogeneity of the liquid beyond the crossover}

In a recent study~\cite{EPL2016}, we found that the heterogeneous nature of the dynamics becomes profound near the thermal crossover, where the self-part of Van Hove correlation $(G_s(r,t))$~\cite{HansenBook06} shows stretched exponential spatial decay ($G_s(r,t) \sim \mathrm{exp}(-lr^k)$).
Here, we extend such studies to track the $N$-dependence of $k(t)$ for different $T$ as shown in  Fig.~\ref{Fig:SVH_Fit500} (panel a: $N=500$, panel b: $N=150$ and panel c: $N=75$) for $V_{\rm conf}^{\rm Ir}$.
Here, we denote $k$ for $V_{\rm conf}^{\rm Ir}$ and $V_{\rm conf}^{\rm Cr}$ as $k_{\rm Ir}$ and $k_{\rm Cr}$, respectively. While the stretched exponential behavior is generic for all $N$, the finer details differ. In particular, the recovery of the final isotropic liquid like behavior signaled by $k_{\rm Ir} \approx 2$ for large $t$ occurs differently for different $N$. In Fig.~\ref{Fig:SVH_Fit500}(c), we see that an isotropic liquid is not at all recovered for $N=75$ up to the largest time of our simulation, even at high $T (=0.10)$. System with $N=500$ particles (Fig.~\ref{Fig:SVH_Fit500}(a)) attains  such limit more smoothly than $N=150$ (Fig.~\ref{Fig:SVH_Fit500}(b)), indicating a surprisingly large temperature window of inhomogeneous liquid (as signaled by $k_{\rm Ir} < 2$) will gradually shrink as our nano-cluster approaches the bulk limit. We find that $k_{\rm Ir}$ attains the minimum value, which increases with $N$, at $T \sim T_X$. Our results for low $T$ indicates that the conjecture of universal exponential tails of $G_s(r,t)$ in glass formers~\cite{Pinaki07} hold even in case of Coulomb clusters.

Similar analysis for circular confinement shows that while $k_{\rm Cr} \approx 1$ at low $T$ , $1<k_{\rm Cr}<2$ at intermediate $T$, for  all $N$, implying a stretched Gaussian decay of $G_s(r,t)$ in contrast to stretched exponential decay observed in irregular confinement (for details see Fig. 4 in supplementary materials). So, we see while the `solid' phase is quite similar in circular and irregular geometry, the Coulomb liquid is quite inhomogeneous in irregular geometry for a wide range of $T (0.02<T<0.25)$. 

\subsection{Temporal bond orientation correlation function}

Recent experimental advances~\cite{Peter2013, Zahn_Maret_PRL2000} show that study of long time dynamical behavior of a system at different $T$ can yield crucial informations in identifying the solid, liquid and even hexatic phases for $2D$ systems. Here, we analyze bond orientational correlation function in time~\cite{Peter2013}, defined as
\begin{equation}
g_{6}(t) = \langle \psi_{6}^*(t) \psi_{6}(0)\rangle
\label{Eq:G6t}
\end{equation} 

In the solid phase $g_{6}(t)$ remains constant in $t$ while it decays exponentially in the isotropic liquid phase for bulk systems. In the hexatic phase, $g_{6}(t)$ shows an algebraic decay, i.e., $g_{6}(t) \sim t^{-\alpha(T)}$ and $\alpha(T) \sim 1/8$~\cite{NelsonBook02, Zahn_Maret_PRL2000} at the hexatic to liquid transition, though the transition turns into a crossover with disorder~\cite{Lyuksyutov}.
\begin{figure}[t]
\includegraphics[width=8.9cm,keepaspectratio]{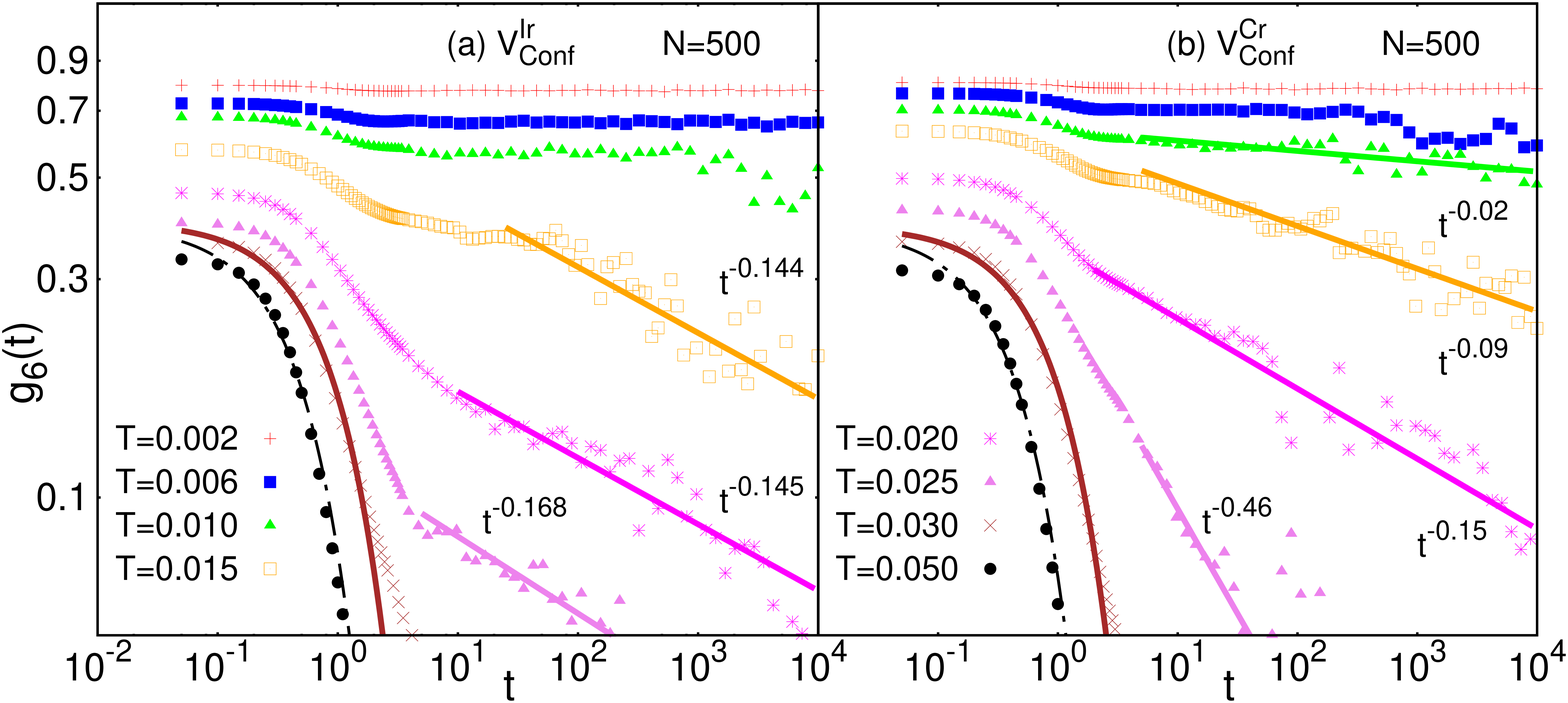}
\caption{
         $t$-dependence of temporal bond orientational correlation function, $g_6(t)$, at different $T$ for irregular (panel a) and circular (panel b) confinements with $N=500$ particles. Solid lines are the appropriate fitting to the actual data (points). The results confirm three qualitatively different evolution: The low-$T$ flat traces typify the solid-like behavior, the power-law decay for intermediate $T$ is reminiscent of the bulk $2D$ hexatic trend, while the large temperature exponential fall represent isotropic liquid nature.
}
\label{Fig:G6t}
\end{figure}

Fig.~\ref{Fig:G6t} shows the $t$ dependence of $g_{6}(t)$ at different $T$ with $N=500$ for IWM (panel a) and CWM (panel b) which clearly identifying three distinct behavior of $g_{6}(t)$ as mentioned: At very low $T$ ($\leq 0.006$), $g_{6}(t)$ remains nearly flat in $t$ implying a `solid-' like phase. It decays exponentially in the high $T (\ge 0.030)$ liquid phase whereas in the intermediate $T$ the decay is algebraic $(g_{6}(t) \sim t^{-\alpha(T)})$. $\alpha(T)$ shows a very weak $T$ dependence in irregular confinement and $\alpha(T_X \simeq 0.02)=0.17$, a value which is  larger than the KTHNY prediction. On the contrary $\alpha(T)$ shows a stronger $T$ dependence in circular confinement with $\alpha(T_X)=0.15$.
Such behavior of $g_6(t)$, qualitatively discerning its power law decay, is a compelling evidence for the existence of a hexatic glass like phase in our irregularly confined Coulomb clusters. Similar inference holds for $N=150$ and is illustrated in Fig.~5 in supplementary materials. 

\subsection{Time scales of structural relaxation}

Signatures of multi-scale temporal relaxations in confined systems is recently elucidated~\cite{EPL2016} in our earlier study.
Here, we focus on extracting relevant time scales for structural relaxation from the overlap function, cage correlation function and persistence and exchange times. We will discuss how these signify the underlying motional processes quantifying the nature of the dynamics. We will finally look into the cross-correlation between these times scales to understand the inter-relations between these time-scales.

\subsubsection{Structural Relaxation Time from Overlap Function}

We estimate the time evolution of the self part of a two-point density correlation function, called the overlap function $Q(t)$~\cite{ SmarajitRev1, SmarajitRev2}, defined as:

\begin{equation}
Q(t) = \left\langle \frac{1}{N} \sum_{i=1}^{N} w(|\vec{r}_i(t_0 + t) - \vec{r}_i(t_0)|)\right\rangle
\end{equation}
where $w(r) = 1.0$ if $r < r_c$ and zero otherwise. The angular parentheses denotes averaging of results over the time origin, $t_0$, and also over different realizations of the disorder. Since particles mostly rattle in a small distance around their equilibrium positions (within $r_c$, as defined below) at small $t$, $Q(t) \sim 1$ where as for large $t, Q(t) \sim 0$ signaling completion of structural relaxation. We evaluate $w(r)$ by monitoring the displacement of individual particles and once a particle moves by a distance greater than $r_c$ then we set $w(r)=0$ for all future times. This definition ensures $Q(t)$ is a monotonically decaying function of $t$.

\begin{figure}[t]
\includegraphics[width=8.8cm,keepaspectratio]{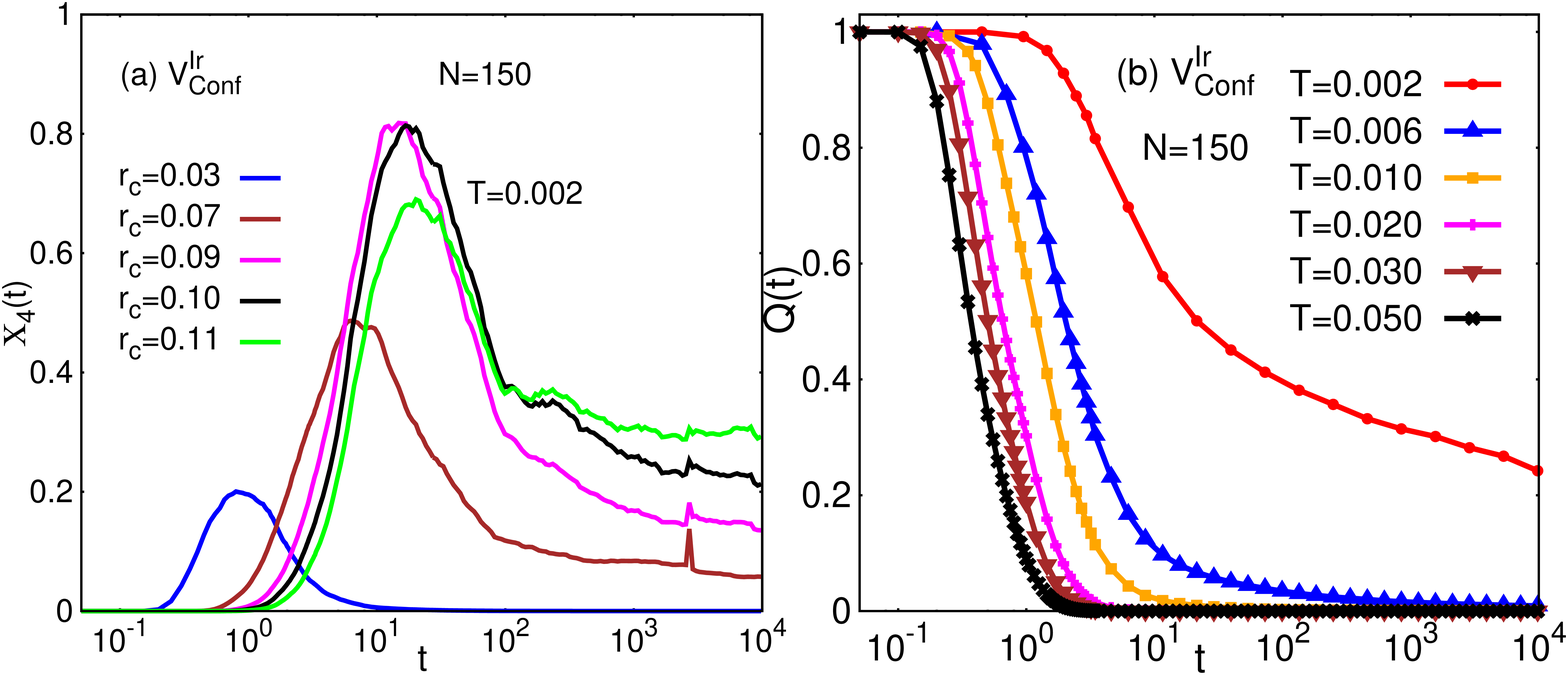}\\
\includegraphics[width=8.9cm,keepaspectratio]{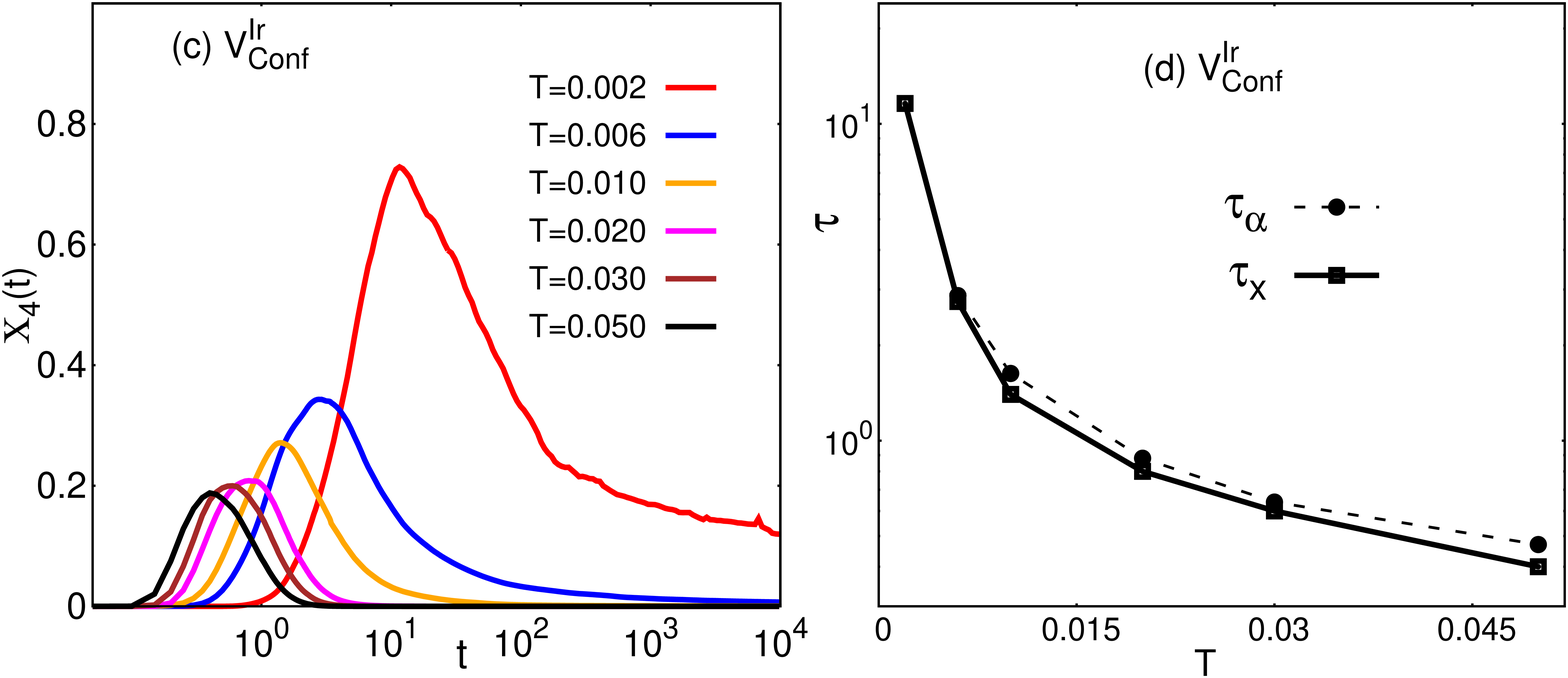}
\caption{
         (a) Determination of the cut-off $r_c$ for $V^{\rm Ir}_{\rm conf}$ with $N=150$, by tracking its value for which $\chi_4(t)$ achieves the maximum keeping temperature fixed at its lowest value ($T=0.002$). This optimal $r_c0.09$ is then used for the evaluation of the overlap function $Q(t)$.
         (b) The time dependence of the overlap function $Q(t)$ for different $T$ for $N=150$. Solid lines are added to the actual data points to guide the eye. While at low $T, Q(t)$ decays very slowly, it falls to zero exponentially for $T >0.006$.
         (c) $t$-dependence of $\chi_4(t)$ at different $T$. $\chi_4(t)$ attains maximum at time $\tau_x$ which decreases with increase in $T$. 
         (d) $T$ dependence of $\tau_x$ along with $\tau_{\alpha}$, obtained from $Q(t)$. Both the time scales show similar $T$ dependence.
}
\label{Fig:Qt_Rc} 
\end{figure}

Obviously, $Q(t)$ depends on the choice of $r_c$ and the optimal $r_c$ is chosen as follows: We study the fluctuation of $Q(t)$, defined as the dynamical susceptibility:
\begin{equation}
\chi_4(t) = \frac{1}{N}[\langle Q^2(t)\rangle - \langle Q(t) \rangle^2]
\end{equation}
at the lowest temperature for different choices of $r_c$. From definition, $\chi_4(t)= 0$ when no particles move by a distance beyond $r_c$. It is zero, as well, when displacements of all particles are greater than $r_c$. So there must be a time scale, $\tau_x$, when $\chi_4(\tau_x)$ reaches maximum for a given $r_c$. Thus, the maximum of $\chi_4$, i.e. the fluctuations in $Q(t)$, at $t=\tau_x$ signify the maximal dynamic heterogeneity. At a given $T$, we choose the value of $r_c$ for which such dynamic heterogeneity attains the maximum. For example, we find that for $N=150$,  $\chi_4(t)$ reaches its maximum for $r_c=0.09$ which is about $14\%$ of $r_0$ (see Fig.~\ref{Fig:Qt_Rc}(a)). At a given $T$, $\chi_4(t)$ attains maximum for a time proportional to the structural relaxation time~\cite{Donati2002,SmarajitRev1}.

Fig.~\ref{Fig:Qt_Rc}(b) shows the $t$-dependence of $Q(t)$ in irregular confinement at different $T$ for $N=150$. While at low $T, Q(t)$ decays very slowly, it falls to zero rapidly for $T >0.006$. The time dependence of $Q(t) \propto \mathrm{exp[-(t/\tau_{\alpha})^c]}$, beyond a very short time, is found exponential $(c=1)$ for $T >0.006$, whereas, it is stretched exponential $(c \approx 0.62)$ for $T=0.006$ and a much slower decay, seemingly power law for $T=0.002$.

To get an estimation of the structural relaxation time, $\tau_{\alpha}$, from the overlap function $Q(t)$, we use $Q(\tau_{\alpha}) = 1/{\mathrm{e}}$ for $T \ge 0.006$. For bulk system, one can estimate the structural relaxation time by looking into the temporal decay of the self part of the intermediate scattering function, $F_s(k,t)$ (with $k$ set to the value at which the static structure factor features its first peak). But, since the wave vector $k$ is not well defined for a finite system, we use $Q(t)$ to calculate $\tau_{\alpha}$.

Next, we study the $t$-dependence of $\chi_4(t)$ at different $T$ (Fig.~\ref{Fig:Qt_Rc}(c)). We find that $\tau_x$ decreases to lower values with increase in $T$. We show the $T$ dependence of $\tau_x$ along with $\tau_{\alpha}$ in Fig.~\ref{Fig:Qt_Rc}(d). The $T$-dependence of these two characteristic times appear to be identical within the tolerance. This implies that the enhanced heterogeneities cause structural relaxation in Coulomb clusters. The rapid increase of $\tau_x$ and $\tau_{\alpha}$ with decrease in $T$ are reminiscent of glassy systems.

\subsubsection{Cage-correlation time}

The dynamics in Coulomb clusters can be further probed by addressing the `cage effects'. In glass formers, the particles get slower, without appreciable change in the structure, upon approaching the glass transition temperature. This is due to the caging effect in which each particle is locked up by its Coulomb repelling neighbours. Rearrangement of `cage' relaxes the system and thereby the particles diffuse. The rate of change of surroundings of a particle yields a cage correlation function (CCF) and helps in understanding how rapidly the local environment of each particle changes on an average.

To calculate CCF we first define a neighbor list keeping track of each particle's  neighbors. If the list of a particle's neighbors at time $t$ is identical to the list at time $t=0$, CCF assumes unity for that particle. If, however, any of the confined particle's neighbor changes, CCF becomes zero at that instant of time. A neighbor list $\mathbf{L_{i}(t)}$ for particle $i$ in an $N$ particle system is a vector of length $N$,  and is defined as~\cite{Khademi}
\begin{equation}
\mathbf{L_{i}(t)}= f(r_{ij})
\label{Eq:Ngbr_lst}
\end{equation}
with $j=1,2,...,N$, and $f(r_{ij})= 1$ if $j$ is the nearest neighbour of $i$ at time $t$ and zero otherwise. We use Voronoi construction to select neighbour list at all $t$. The CCF at $t$ is given by
\begin{equation}
C_g(t)=\frac{\langle \mathbf{L_{i}(t)} \cdot \mathbf{L_{i}(0)} \rangle}{\langle \mathbf{L_{i}^2(0)}\rangle} 
\label{Eq:CageCorr}
\end{equation}

\begin{figure}[t]
\includegraphics[width=9.0cm,keepaspectratio]{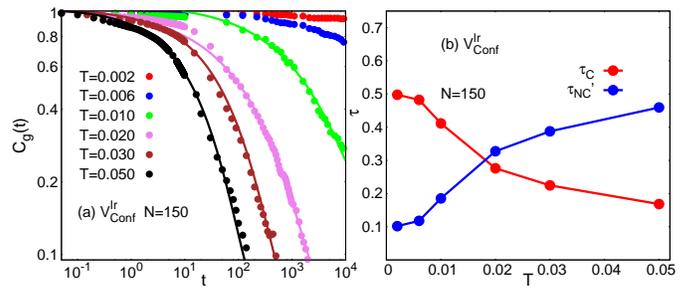}
\caption{
         (a) Decay of cage correlation function, $c_g(t)$, at different $T$ for $N=150$ depicting how the initial neighbors fail, with time, to confine a particle, on the average. The solidity at low-$T$ is reflected in the fact that the same neighbors continue to cage a particle at all time. (b) $T$-dependence of average caging $(\tau_c)$ and non-caging $(\tau_{NC})$ time for $N=150$, illustrating the thermal crossover at $T \sim T_X$.
}
\label{Fig:Cgt}
\end{figure}

In Fig.~\ref{Fig:Cgt}(a) we show the decay of CCF at different $T$ for irregular confinement. We see that at $T=0.002$, $C_g(t)$ remains close to unity for all times, implying no significant rearrangements of neighbours. This is because, particles are just jiggling only around their equilibrium positions. As $T$ increases $C_g(t)$ shows a decay that turn out to stretched exponential in $t$: $c_g(t) \propto \mathrm{exp} \left[ -t/\tau_g \right]^{\beta} $ with $\beta <1$ (details of the fitting parameters are given in Table I in supplementary materials). Such decay is found for both the confinements (for circular confinement see Fig.~7 in supplementary material). Similar decay is also common to supercooled liquids~\cite{Rabani}. Recently, it is found that water in nanoporous silica~\cite{Nano_cage} shows stretched exponential decay with the characteristics exponent $\beta =0.6$. Ref.~\cite{Phillips} describing stretched exponential temporal decay in supercooled liquids with $\beta \approx 0.6$. While the literature discusses only the short-range interacting systems, it is interesting to note that our Coulomb clusters (with long range interactions) show similar trends. For a system with with static random traps, the asymptotic value of $\beta$ relates the dimensionality $d$ of the system through $\beta = d/(d + 2)$~\cite{Cage_Corr_beta}. Thus for $d=2$, asymptotic value of $\beta$ is $0.5$. We find that for both the circular and irregular confinements $0.45<\beta<0.6$ which we consider is in broad agreement with such predictions.

We estimate the caging and non-caging times by studying the relative fluctuations in position of the particles with respect to its neighbors. A perfectly caged particle is expected to have equidistant neighbors. Exploiting this idea, we define the caging time $(\tau_{C})$ of the $i$-th  particle as the time up to which the following condition holds for at least three of its neighboring particles $(j)$:
\[
\frac{\sigma_i(t)}{\Delta r_{ij}(t)} \le d;~~~\text{ for at least}~~ j=3
\]
Here $j=1,2, \cdots N_b$, $N_b$ denoting the number of nearest neighbors of particle $i$; $\Delta r_{ij}(t) = | \vec{r}_i(t) - \vec{r}_j(t)|$ and $\sigma^2_i(t) =\langle \Delta^2 r_{ij}(t) \rangle -  \langle \Delta r_{ij}(t) \rangle^2$. We take cut-off value $d$ as $0.1 r_0$ (analogous to the usual Lindemann ratio). Once the above criterion breaks down, then the time taken by the particle to get caged again, is called the non-caging time $(\tau_{NC})$. In above, caging is defined at least with respect to three of its neighbors, because in $\mathrm{2D}$ a minimum of three particles are required to `cage' a particle. 
Fig.~\ref{Fig:Cgt}(b) shows the averaged $\tau_{C}$ and $\tau_{NC}$ for irregular confinements. Here, $\tau_{C}$ and $\tau_{NC}$ are expressed relative to the total time of the MD simulation. At low $T$, caging time is expectedly high as particles are mostly confined by its nearest neighbors. As $T$ increases thermal energy overcomes caging. Thus, $\tau_{C}$ decreases with  $T$ while $\tau_{NC}$ increases. From  Fig.~~\ref{Fig:Cgt}(b) we see that these two time scales crosses each other around the same $T_X$ that shows other signatures of the thermal crossover. We note that this analysis excludes the particles on the boundary because the current definition of `caging' is ill defined for them.

\subsubsection{Persistence and exchange time}

A crucial time scale that probes the relaxation mechanism as well as the glassiness of dynamics is the behavior of the persistence and exchange times~\cite{Chandler,PstExc}. They are defined as follows: At any given $T$, let us consider a particle $i$, whose initial $(t=0)$ position position is $r_i (0)$. The persistence time $t_1$ for a given cut-off distance $d$ is specified by the first time that particle $i$ has moved far enough that $|r_i(t_1)-r_i(0)| \ge r_p$.
\begin{figure}[t]
\includegraphics[width=8.7cm,keepaspectratio]{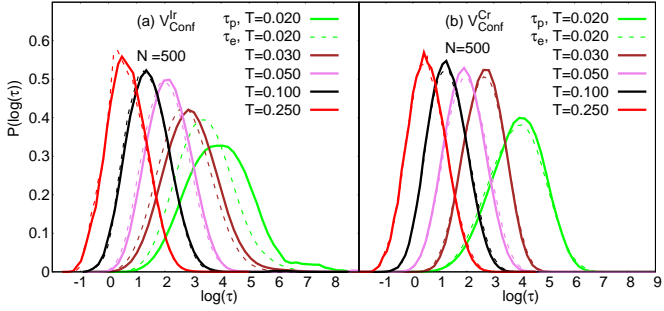}
\caption{
         Distribution of persistence $(\tau_p)$ and exchange $(\tau_e)$ times for (a) irregular and (b) circular confinements with $N=500$ particles. The decoupling of the two distributions at low $T$ demonstrates the qualitative similarity of the particle dynamics in irregular traps with those in glassy systems. Such decoupling is hard to discern for circular traps.
}
\label{Fig:PsExtm}
\end{figure}

Exchange times $t_n - t_{n-1}$ for $n >1$, require the recursive determination $t_n$, setting $t_{n-1}$ as initial time. For a given $r_p$, distributions of exchange $(\tau_e)$ and persistence $(\tau_p)$ times can be obtained by ensemble averaging over all trajectories. $r_p$ is chosen of the order of $r_0$ so as to probe the structural relaxation and diffusion. The same $r_p$ is used for all $T$. It has been shown~\cite{Chandler,PstExc} that for glassy dynamics, distribution of $(\tau_e)$ and $(\tau_p)$ generally decouple near the glass transition temperature. This decoupling is attributed to dynamic heterogeneity in these systems, and constitute a key feature of supercooled liquids~\cite{Chandler,PstExc}.

The distribution of these two time scales are shown in Fig.~\ref{Fig:PsExtm} for irregular (panel a) and circular (panel b) confinements with $N=500$. We notice that for $T>0.030$, $P(\tau_e)$ and $P(\tau_p)$ coincide for both the confinements. As $T$ becomes close to $T_X (\approx 0.020)$, the two distributions become distinct; mean value for $P(\tau_p)$ moves toward longer times for IWM. In contrast, for circular confinement distribution of these two time scales do not appear to decouple. We have ensured that the decoupling is not because of bias in the sampling process by carrying out Jackknife test~\cite{Num_Recipe}. Such decoupling is also observed for $N=150$ (see Fig. 8 in supplementary materials). Note that the distributions of $\tau_e$ and $\tau_p$ are shown for $T \ge T_X$ as at low $T$, displacement of particles beyond $r_p$ is rare and thus such distribution lacks confidence below $T_X$.

\begin{figure}[t]
\includegraphics[width=8.5cm,keepaspectratio]{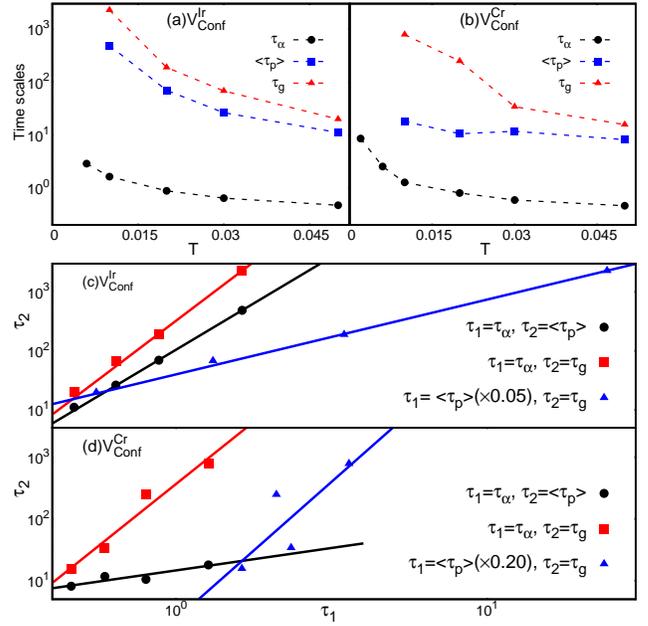}
\caption{
         $T$-dependence of structural relaxation time $\tau_{\alpha}$, average persistence time $\langle \tau_p \rangle$ and cage correlation time $\tau_g$ for (a) irregular and (b) circular confinements. Dotted lines are for visual guidance. Panel (c-d): Thick dots show the cross-Correlation between $\tau_{\alpha}, \langle \tau_p \rangle$, and $\tau_g$ in a log-log plot for irregular and circular confinement, respectively, taken over a temperature range $0.01 \leq T \leq 0.05$. Solid lines represent the best fit to such correlations. $\langle \tau_p \rangle$ is scaled by appropriate factors for visual clarity.
}
\label{Fig:TimeScales}
\end{figure}

Since, the overlap function, cage correlation function and persistence time, all are connected with the structural relaxation of the system, we now look into the $T$-dependence of the characteristic time scales associated with these quantities.For example, $\tau_g$ (obtained by fitting the cage correlation function $c_g(t)$ as discussed) represents characteristic time scale for the rearrangement of particles in their local environment and average persistence time, $\langle \tau_p \rangle$, depicts the time required for particles to move a certain distance on average and thus associated with the relaxation mechanism in the system. In Fig.~\ref{Fig:TimeScales}, we show the $T$-dependence of $\tau_{\alpha}, \tau_g$ and $\langle \tau_p \rangle$ for irregular (panel a) and circular confinements (panel b). We find that all these time scales increase rapidly for IWM compared to CWM as $T$ approaches $T_X$. Thus, the signatures of glassy dynamics, while show up in both confinements, are more prominent for irregular confinement, broadly mimicking a bulk disorder system. 

\vspace*{0.5cm}
\subsubsection{Cross-Correlation between different relaxation times}

Here, we study the cross-correlation between the different time scales described above. Fig.~\ref{Fig:TimeScales}(c) illustrates the cross-correlation between $\tau_{\alpha}$ and $\tau_g$, $\tau_{\alpha}$ and $\langle \tau_p \rangle$, and $\langle \tau_p \rangle$  and $\tau_g$, gathered over a temperature window across melting; namely, $0.01 \le T \le 0.05$, for irregular confinement. Our results indicate that there is a positive correlation between these characteristic times scale; when one increases, other also increases in a proportionate manner.
Quantifying such correlations, we find that $\tau_i \propto \tau_j^{c_{ij}}$ with following sets of \{$\tau_i, \tau_j, c_{ij}$\}: \{$\tau_{\alpha}, \tau_p, 3.14 \}$, $\{\tau_{\alpha}, \tau_g, 3.98 \}$ and $\{\tau_p, \tau_g, 1.27 \}$. Similar positive correlation between different times scales is also observed for circular confinement as depicted in Fig.~\ref{Fig:TimeScales}(d). In this case, we find \{$\tau_i, \tau_j, c_{ij}$\}: $\{\tau_{\alpha}, \tau_p, 0.73 \}$, $\{\tau_{\alpha}, \tau_g, 4.03 \}$ and $\{\tau_p, \tau_g, 4.45 \}$. Thus, all these time scales give a coherent description of the nature of the dynamics of particles in traps.

\section{CONCLUSION}

In conclusion, we have carried out comprehensive simulations to study the thermal crossover for Coulomb-particles in confinements, by probing their static and dynamic properties. While static properties are found quite insensitive to the symmetry of the confinements, dynamical correlations yield distinct signatures for the two confinements. We find intriguing stretched exponential spatial correlations in irregular confinements and a stretched Gaussian behavior in circular traps. Such a  stretched Gaussian trend is recently observed in experiments with nano-particles in confined media~\cite{Nanopost14,Nanopost15}. We devised several melting criteria, often taking advantage of the confined geometries, all of which point towards a single crossover temperature, when expressed in term of bulk-parameter $\Gamma$, yields $\Gamma \approx 147$ and and is quite close to the predicted critical value, $\Gamma_c \approx 137$, for the bulk system. 

Our results offer insights into the `phases' in a confined and finite system. From the study of static correlations, we find that the low-$T$ phase lacks positional order but posses bond-orientational order, which is further strengthened by temporal bond orientation correlation function showing an algebraic decay below the cross over temperature. We also witness key signatures of glassy behavior in our systems, such as, slow heterogeneous dynamics, rapid increase of relaxation time, decoupling of persistence and exchange times, etc. All these indicate that a hexatic-glass like phase is possible to realize in finite systems, at least in those with irregularity and perhaps for long-range inter-particle interactions. A hexatic glass, differing from the usual hexatic phase in $\mathrm{2D}$ only by its quenched defects -- makes the study of defects in our system an interesting proposition. We hope that all these will help enriching our knowledge of thermal melting in two-dimension.

\section*{ACKNOWLEDGEMENTS}

The authors acknowledge computational facilities at IISER Kolkata. BA acknowledges  University  Grant  Commission (UGC), India, for doctoral fellowship. AG acknowledges the hospitality of the Aspen Center for Physics, and was supported in part by the Simons Foundation.

\bibliography{BA_Draft_Supple}
\bibliographystyle{apsrev}

\pagebreak
\widetext
\clearpage 
~\vspace{2cm} 

\setcounter{section}{0}
\setcounter{equation}{0}
\setcounter{figure}{0}
\setcounter{table}{0}
\setcounter{page}{1}
\makeatletter
\renewcommand{\theequation}{S\arabic{equation}}
\renewcommand{\thesection}{\Roman{section}} 
\renewcommand{\thefigure}{S\arabic{figure}}
\renewcommand{\bibnumfmt}[1]{[S#1]}
\renewcommand{\citenumfont}[1]{S#1} 

\begin{center}
{\Large\bf Supplementary materials: Static and Dynamic Properties of Two Dimensional Coulomb Clusters}
\end{center}

\vspace{0.5cm}

\begin{center}
\vspace{0.5cm}

Biswarup Ash\textsuperscript{1}, J. Chakrabarti\textsuperscript{2}, and Amit Ghosal\textsuperscript{1}

\vspace{0.5cm}
{\em \textsuperscript{1}Indian Institute of Science Education and Research-Kolkata, Mohanpur, India-741246.

     \vspace{0.3cm}
     \textsuperscript{2}S.N. Bose National Centre for Basic Sciences, Block-JD, Sector-III, Salt Lake, Kolkata-700098.}\\[15mm]

\end{center}

In this supplementary material, we discuss results related to Coulomb interacting particles in circular confinement, $V_{\rm conf}^{\rm Cr}$, for a comparison with the corresponding results, discussed in the main article, on irregular confinement.

\section{Positional order in circular confinement}

In this section, we illustrate and quantify the positional order in circular confinement with Coulomb interacting particles for a fair comparison with similar study in Fig.~1 of main article. The results in Fig.~\ref{fig:CWM_FT}(b) indicates that, in the ground state $(T=0)$, the positional order in $V_{\rm conf}^{\rm Cr}$ is better developed than in $V_{\rm conf}^{\rm Ir}$ (see Fig. 1 of main paper), producing more definitive peaks in the Fourier transform of density. This is even more prominent (Fig.~\ref{fig:CWM_FT}(c)) in the subsystem of selected particles where the stronger order is evident in naked eye (Fig.~\ref{fig:CWM_FT}(a)). At $T=0.020$, $\rho(\vec{k})$ shows a diffuse pattern  (Fig.~\ref{fig:CWM_FT}(d)) similar to that observed for $V_{\rm conf}^{\rm Ir}$ (see Fig. 1(d)).

\renewcommand{\thefigure}{S.\arabic{figure}}
\begin{figure}[h!]
\includegraphics[width=10cm,keepaspectratio]{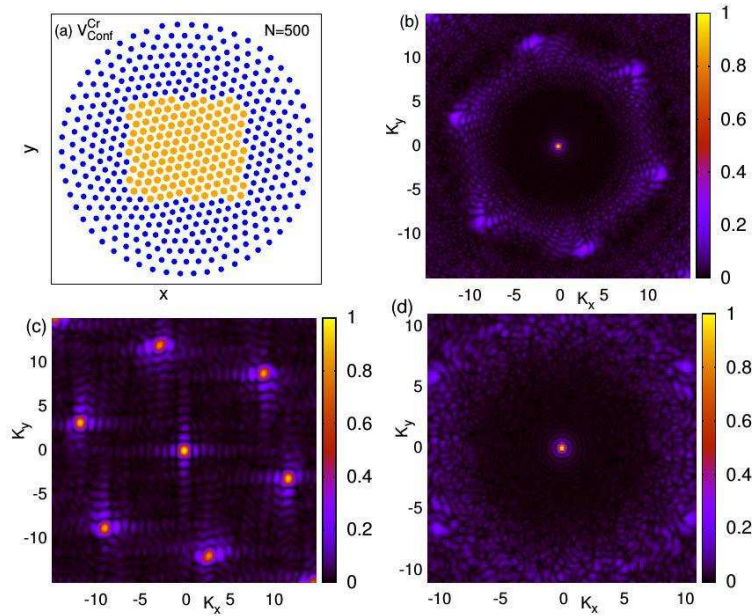}
\caption{
	  (a) The ground state ($T=0$) configuration of $N=500$ particles in circular confinement. The orange dots are particles in the central region selected to search for positional order, while all other particles are represented by blue dots. Panel (b)-(d) show Fourier transform of particle density for various cases: Panel (b) is for the ground state with all particles shown in panel (a). Panel (c) is for the particles only in the central region of panel (a). Panel (d) is for all particles, but at $T=0.02 \sim T_X$. In panel (b-d) magnitude of
$\rho(\vec{k})$ is scaled to unity for visual clarity.
}
\label{fig:CWM_FT}
\end{figure}

\section{Bond Orientational Order for circular confinement}

\begin{figure}[t]
\includegraphics[width=8.5cm,keepaspectratio]{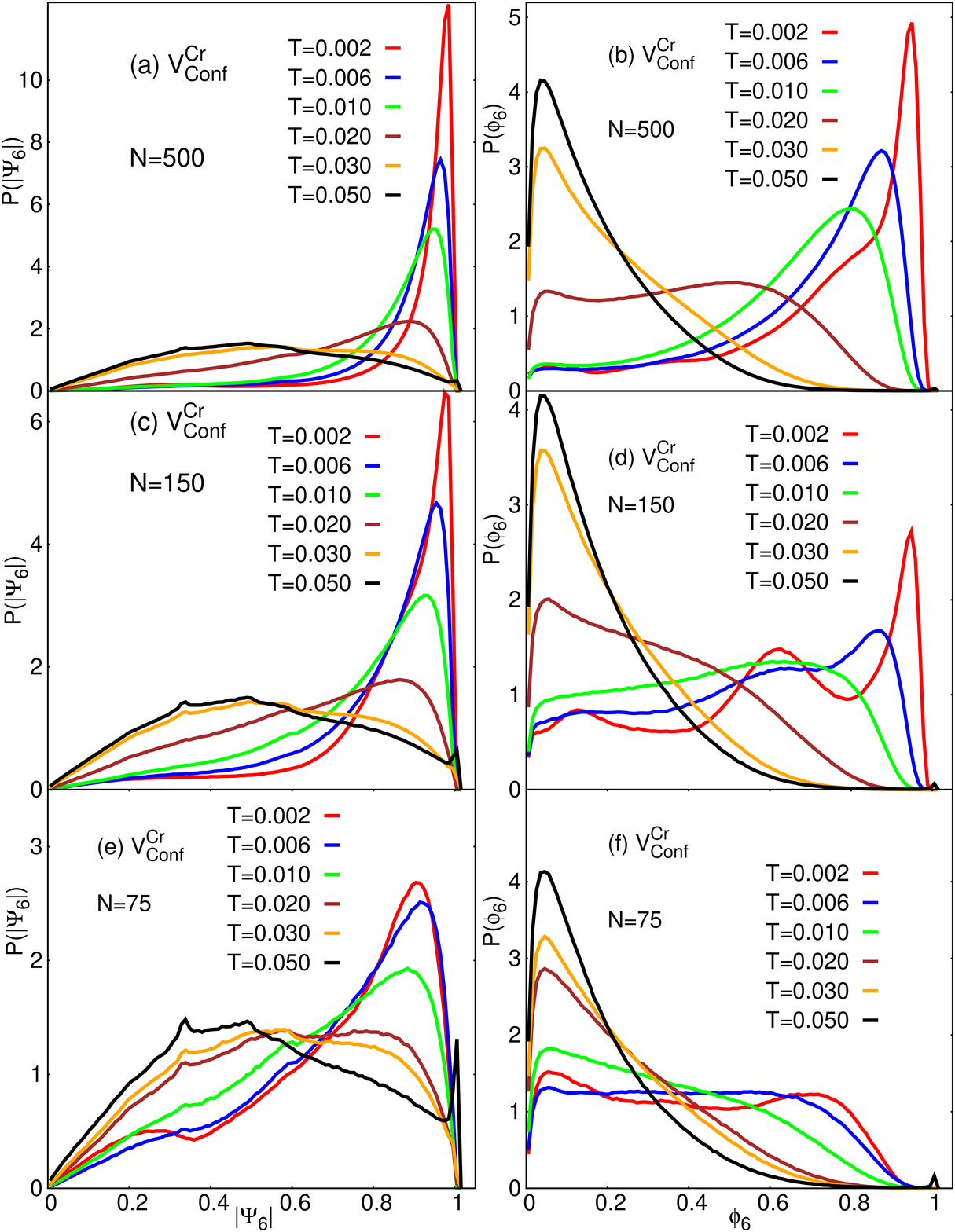}
\caption{
         $T-$dependence of the bond orientational in circular confinement is shown for $N=500$ in panel (a), for $N=150$ in panel (c) and for $N=75$ in panel (e). The low-$T$ peak in $P(|\Psi_6|)$ at $|\Psi_6| \sim 1$ smears out as the system undergoes the thermal crossover. The small peak in $P(|\Psi_6|)$ at $|\Psi_6| \sim 1$ at high $T$, particularly for smaller $N$, does not constitute BOO.
Similar distribution of $\phi_6$ representing the bond orientational correlation up to nearest neighboring distance are shown in panels (b, d, f). $P(\phi_6)$ features a bimodal structures, its peak near $\phi_6 \sim 1$ (at low $T$) signaling solidity and the one at $\phi_6 \approx 0$ describing liquidity.
 }
\label{fig:CWM_psi61}
\end{figure}

Fig.~\ref{fig:CWM_psi61} shows the $T$-dependence of the distribution $P(|\Psi_6|)$ of bond orientational order parameter, $\left\vert \psi_{6} \right\vert$, for $V_{\rm conf}^{\rm Cr}$ with $N=500$ (panel a), $150$ (panel c) and $75$ (panel e) particles. $P(|\Psi_6|)$ shows a sharp peak around unity, depicting an orientationally ordered phase at low $T$, and the distribution becomes broader with increase in $T$. In Fig.~\ref{fig:CWM_psi61}(b,e and f) we show the distribution $P(\phi_{6})$ of $\phi_{6}$ (see main text for definition), at different $T$ for $V_{\rm conf}^{\rm Cr}$ with $N=500, 150$ and $75$, respectively. Here, we find that $P(\phi_{6})$ distinguishes the ordered solid-like phase at low-$T$ and orientationally disordered liquid-like phase at high-$T$. Thus, the temperature dependence of $P(|\Psi_6|)$ and $P(\phi_{6})$ for $V_{\rm conf}^{\rm Cr}$ are qualitatively similar to that discussed for $V_{\rm conf}^{\rm Ir}$ (see Fig. 2) in main paper. However, details vary with respect to $P(\phi_{6})$ for smaller systems (Fig.~\ref{fig:CWM_psi61}(d,f)). $V_{\rm conf}^{\rm Cr}$ seems to present a gradual crossover for $N \le 150$.

\section{Analysis of trajectory of particles in circular confinement}

\begin{figure}
\includegraphics[width=8.5cm,keepaspectratio]{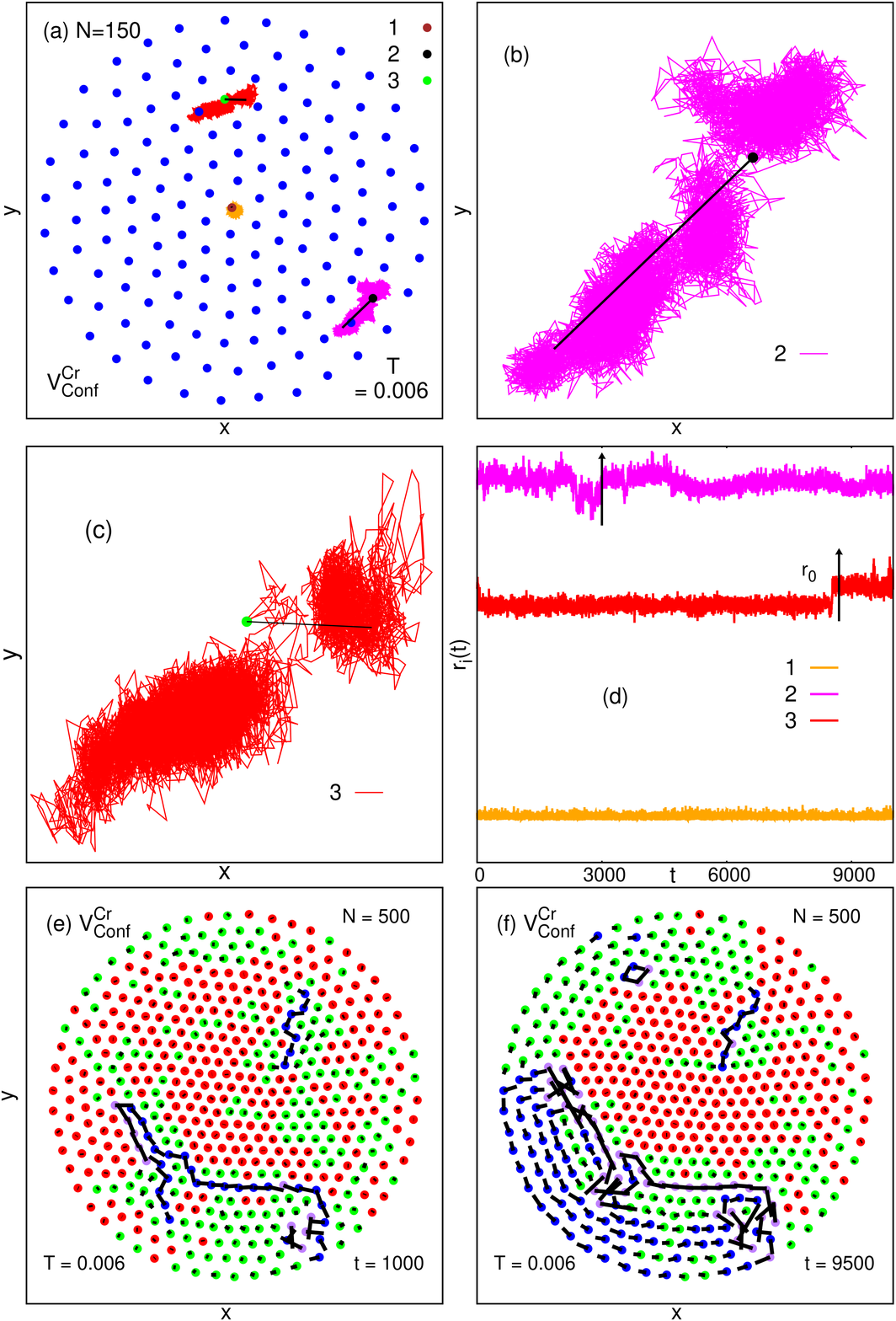}
\caption{
	(a) Initial positions of $N=150$ particles (blue dots) in circular confinement along with the trajectories of three particles at $T=0.006$. Solid lines indicate the net displacement of the particles over the duration of the simulation. (b-c) Illustrates the trajectories of the particles $2$ and $3$ with larger displacements as shown in (a). Both of them jiggles around their equilibrium position for a while and they make occasional and sudden jumps by distance $\sim r_0$, the average inter-particle distance. Particle $1$, however, does not make such hops (not shown separately). (d) $t$-dependence of the instantaneous distance $r_i(t)$ from the centre of the confinement for the three selected particles $(i=1,2,3)$. Displacement ($\Delta \vec{r}_i(t) = \{\vec{r}_i(t) - \vec{r}_i(0)\}$) of $N=500$ particles in circular confinement is shown for (e) $t=1000$ and (f) $ t=9500$ at $T=0.006$.
 }
\label{fig:CWM_Disp}
\end{figure}

Fig.~\ref{fig:CWM_Disp} shows the trajectory of particles in circular confinement while corresponding analysis for irregular confinement is shown in Fig. 4 of the main paper. Motional features are qualitatively similar for particles in both the traps, but here, we emphasise following  two points: \\
(i)  Analysis of the instantaneous position $r_i(t)$ (Fig.~\ref{fig:CWM_Disp}(d)) in $V_{\rm conf}^{\rm Cr}$ reveals that particles change their positions more gradually compared to those in $V_{\rm conf}^{\rm Ir}$ (see Fig. 4).\\
(ii) Circular symmetry dictates the development of coherent motion in $V_{\rm conf}^{\rm Cr}$. Lack of such spatial symmetry in case of $V_{\rm conf}^{\rm Ir}$ causes formation of tortuous string-like path through out the system (see Fig. 4).

\section{Time evolution of exponent $k_{\rm Cr}$}

\begin{figure}
\includegraphics[width=8.5cm,keepaspectratio]{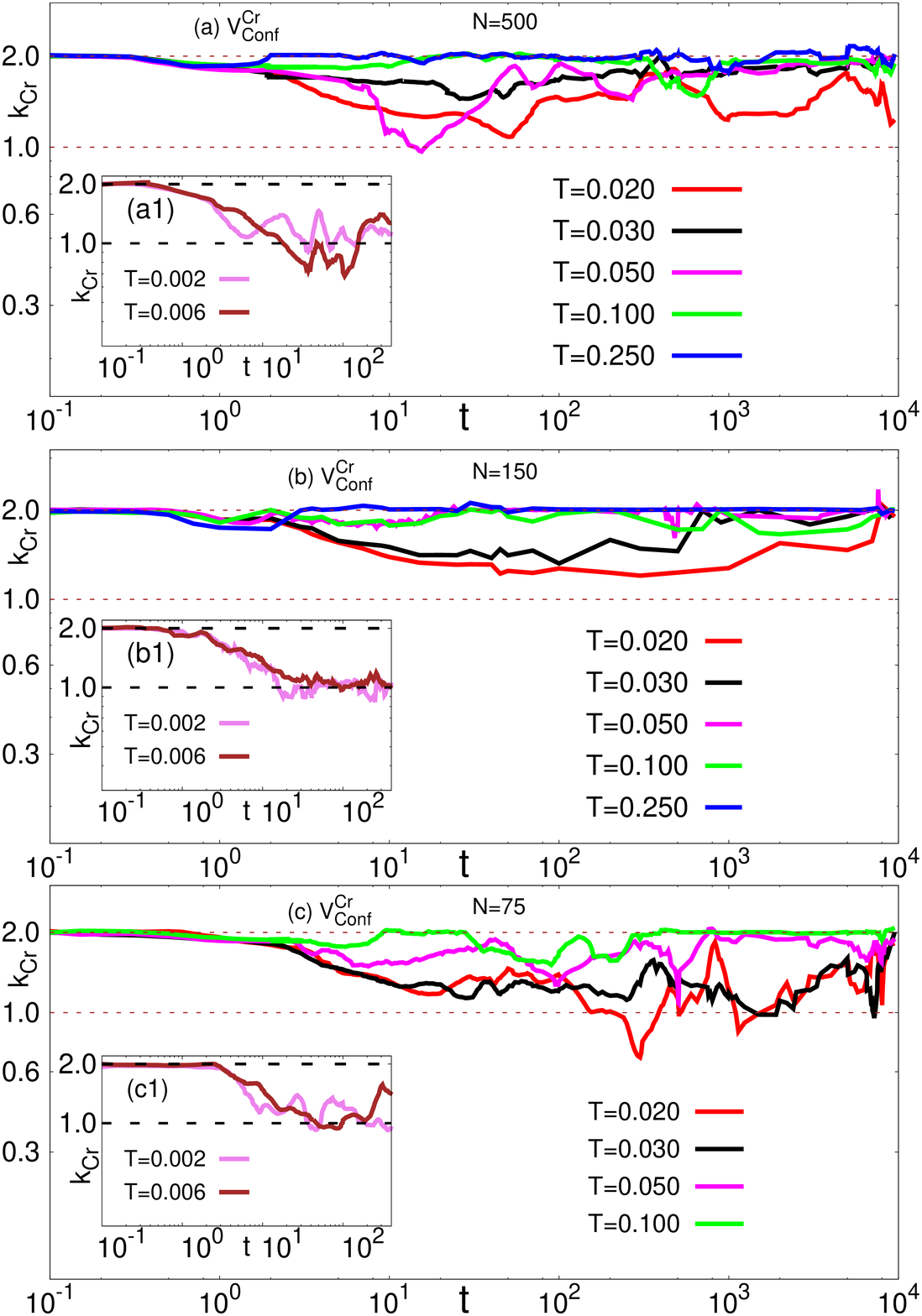}
\caption{
         $t$-dependence of the exponent $\mathrm{k_{Cr}}$ depicting the nature of the spatial evolution of the displacement of the particles, at different $T$ for circular confinement is analyzed for (a)$N=500$, (b) $N=150$ and (c) $N=75$. 
 }
\label{fig:CWM_K}
\end{figure}

In Fig.~\ref{fig:CWM_K} we show the time evolution of the exponent $k_{\rm Cr}$ (see main text for details), at different $T$ for $V_{\rm conf}^{\rm Cr}$ with (a)$N=500$, (b) $N=150$ and (c) $N=75$ particles. While at low $T ( \approx 0.006)$ $k_{\rm Cr}$ continuously decays from $2$ to nearly $1$ (inset (a1), (b1) and (c1)) for all system sizes, $1 \leq k_{\rm Cr} \leq 2$ for all higher $T$. Thus, the distribution of the displacement of the particles is stretched Gaussian for $V_{\rm conf}^{\rm Cr}$ while it is stretched exponential $(k_{\rm Ir} <1 )$ for $V_{\rm conf}^{\rm Ir}$ (see Fig. 5) as discussed in the main article. Interestingly, stretched Gaussian spatial decay has been realized in recent experiments with nano-particles (see Ref. 80-81 in the main paper). Note that for $N=75$, $1 \leq k_{\rm Cr} \leq 2$ but $k_{\rm Ir}<1$ (see Fig. 5(c)) for $T \ge 0.02$. Thus, standard liquid like behavior is not recovered for $V_{\rm conf}^{\rm Ir}$ with small number of particles.

\section{Temporal bond orientation correlation function for Circular confinement}

\begin{figure}[t!]
\includegraphics[width=10.0cm,keepaspectratio]{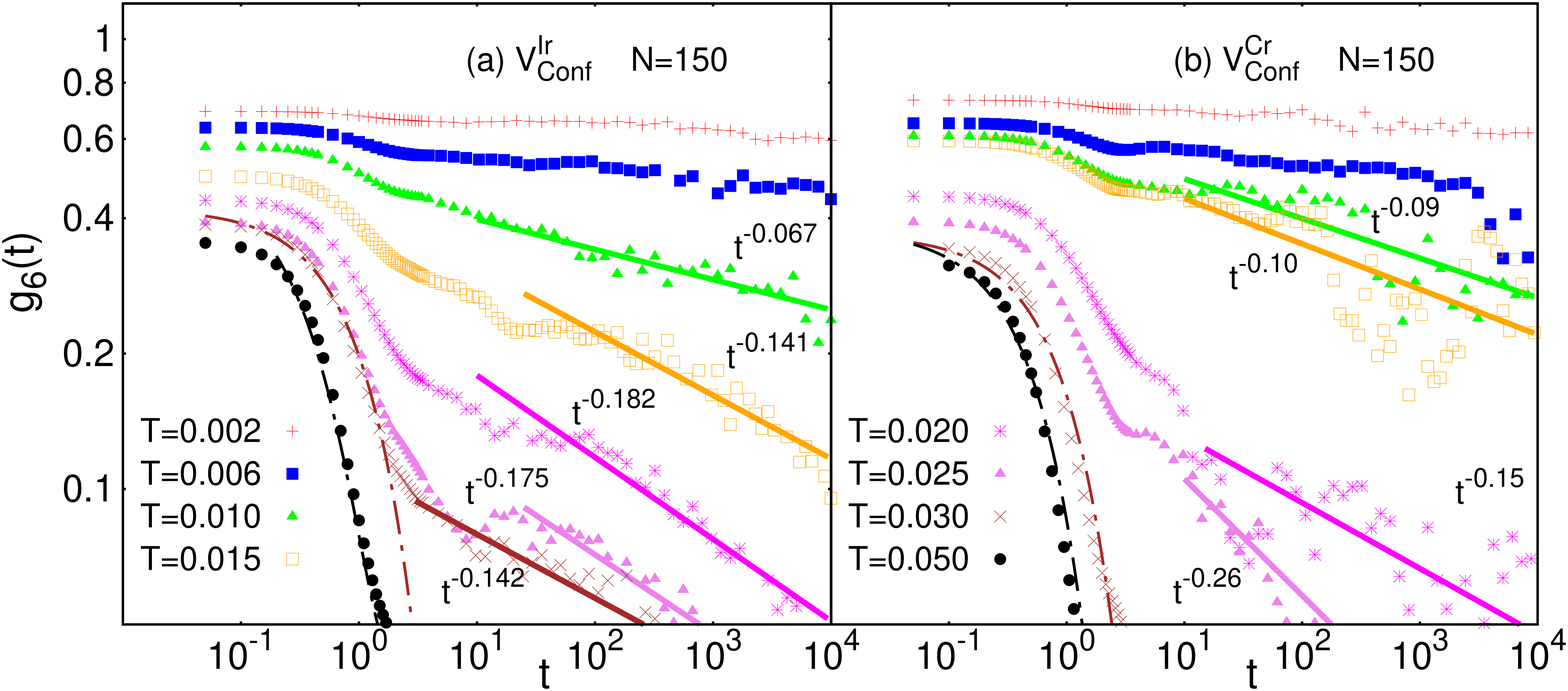}
\caption{
	  $t$-dependence of temporal bond orientational correlation function $(g_6(t))$ at different $T$ for irregular (panel a) and circular (panel b) confinements with $N=150$ particles. Solid lines are the appropriate fitting to the actual data (points). At low-$T$ flat $g_6(t)$ signify the solid-like behavior, the power-law decay for intermediate $T$ is reminiscent of the bulk $2D$ hexatic trend, while exponential fall at large $T$ represent isotropic liquid nature.
 }
\label{fig:CWM_g6t}
\end{figure}

While Fig. 6 in the main article shows the $t$-dependence of temporal bond orientational correlation function $(g_6(t))$ at different $T$ for $N=500$ particles, here in Fig.~\ref{fig:CWM_g6t} we discuss the time evolution of $g_6(t)$ for irregular (panel a) and circular (panel b) confinements with $N=150$ particles, respectively. For $N=150$, the $t$-dependence of $g_6(t)$ at different $T$ for both the traps is  qualitatively similar to what is discussed in the main article. Here, we find that the algebraic decay of $g_6(t)$ persists over wider temperature window for $N=150$ (Fig.~\ref{fig:CWM_g6t}(a)) compared to $N=500$ particles in $V_{\rm conf}^{\rm Ir}$ (see Fig. 6(a)). At intermediate $T$, data show larger fluctuations at long time for both the confinements due to smaller system size.

\section{Overlap function $Q(t)$ for circular confinement }

\begin{figure}[h!]
\includegraphics[width=9cm,keepaspectratio]{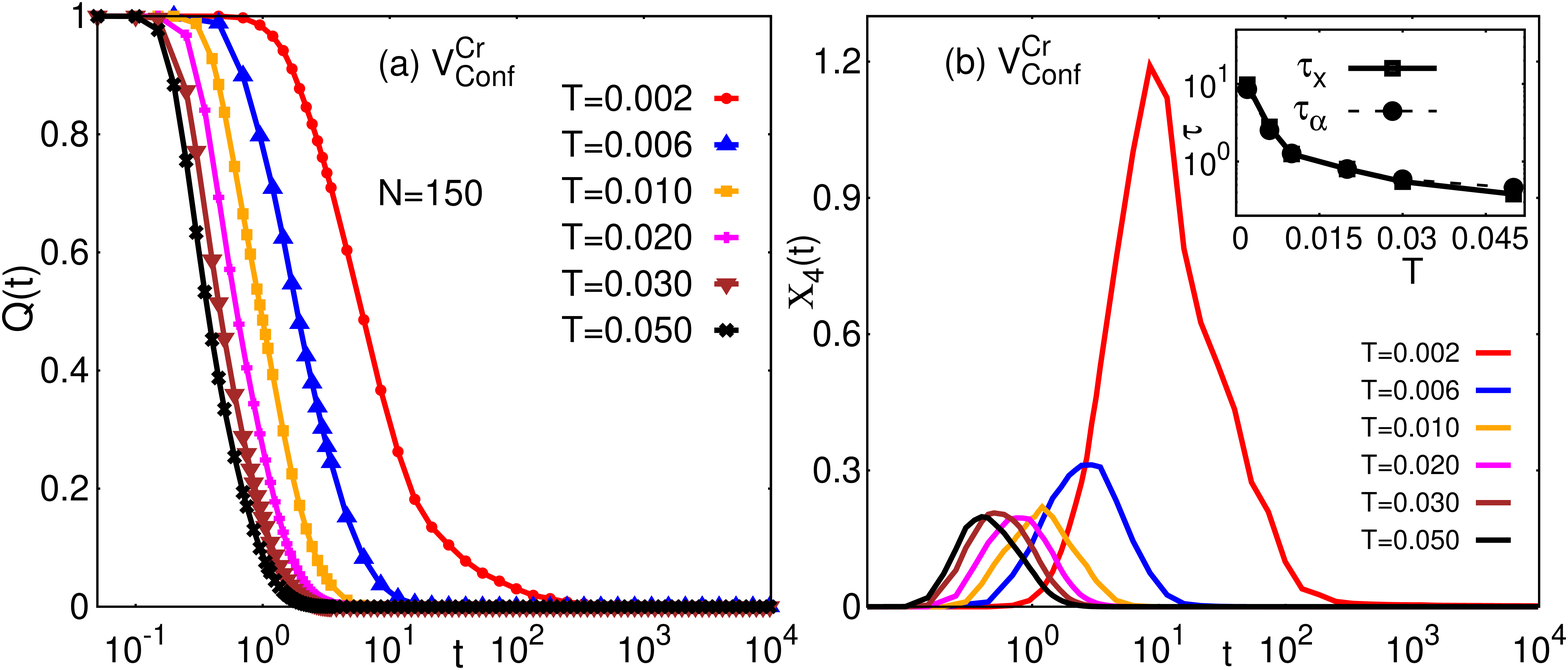}
\caption{
	 (a) Time dependence of the overlap function $Q(t)$ at different $T$ for $N=150$ particles in circular confinement. At $T=0.002$, the decay of $Q(t)$ is stretched exponential while at higher $T$ it is exponential. (b) $t-$dependence of $\chi_4(t)$ at different $T$ for circular confinement with $N=150$. $\chi_4(t)$ attains maximum at time $\tau_x$ which decreases with increase in $T$. Inset shows the $T$ dependence of $\tau_x$ along with $\tau_{\alpha}$, obtained from $Q(t)$. Both the time scales show similar $T$ dependence.
 }
\label{fig:CWM_Qt}
\end{figure}

Fig.~\ref{fig:CWM_Qt}(a) shows the $t$-dependence of the overlap function $Q(t)$ at different $T$ for $N=150$ particles in $V_{\rm conf}^{\rm Cr}$. At low $T$, the decay of $Q(t)$ is quite faster than that in irregular confinement (Fig. 7(b)) and shows stretched exponential decay. At high $T$, Q(t) decays exponentially to zero. Fig.~\ref{fig:CWM_Qt}(b) shows $t-$dependence of $\chi_4(t)$ at different $T$. Here, we see that with increase in $T$, the time, $\tau_x$, at which $\chi_4(t)$ attains the maximum value decreases gradually (inset of Fig.~\ref{fig:CWM_Qt}(b)). As discussed in the main paper, we have extracted the structural relaxation time $\tau_{\alpha}$ from the condition $Q(\tau_{\alpha})=1/e$. The $T$-dependence of $\tau_{\alpha}$ and $\tau_x$ are shown in the inset of Fig.~\ref{fig:CWM_Qt}(b). While both the time scales show similar $T$ dependence, but for $V_{\rm conf}^{\rm Ir}$ these two time scales increase more rapidly with decrease in $T$ (see Fig. 7(d)). Thus, $T$-dependence of structural relaxation time indicates `better glassiness' in $V_{\rm conf}^{\rm Ir}$. 

\section{Details of the fitting parameters for Cage correlation function}

The cage correlation function, $c_g(t)$, is fitted with the following functional form:
\[
c_g(t) \propto \mathrm{exp} \left( -\left(\frac{t}{\tau_g}\right)^{\beta} \right)
\]
For both the confinements, $c_g(t)$ shows stretched exponential decay $(\beta <1)$ and values of $\tau_g$ and $\beta$ at different $T$ is given in Table.~\ref{Table:CagrCorr}. Here, we find that $\tau_g$, the characteristic time scale  associated with the local rearrangement of particles, grows more rapidly with decrease in $T$, for irregular confinement, reinstating the glassy dynamics in $V_{\rm conf}^{\rm Ir}$.
\begin{table}[h!]
\centering
\renewcommand{\arraystretch}{1.5}
\begin{tabular*}{\columnwidth}{@{\extracolsep{\fill}}|l|l|l|l|l|@{}}\hline
\multicolumn{1}{|c|}{Temperature} &
\multicolumn{2}{c|}{Irregular}    &
\multicolumn{2}{c|}{Circular}\\ \hline
      & $\tau_g$  & $\beta$  & $\tau_g$ & $\beta$ \\ \hline
0.010 & 2271.83   & 0.446    & 783.26   & 0.508   \\ \hline
0.020 & 188.002   & 0.442    & 247.56   & 0.574   \\ \hline
0.030 & 67.6891   & 0.553    & 34.035   & 0.639   \\ \hline
0.050 & 19.8172   & 0.596    & 15.638   & 0.603   \\ \hline
\end{tabular*}
\caption{Details of the fitting parameters, $\tau_g$ and $\beta$, of the cage correlation function, $c_g(t)$, for circular and irregular confinements. For both the confinements the decay of $c_g(t)$ is stretched exponential.}
\label{Table:CagrCorr}
\end{table}

\section{Cage correlation function, caging and non-caging time}

\begin{figure}
\includegraphics[width=9cm,keepaspectratio]{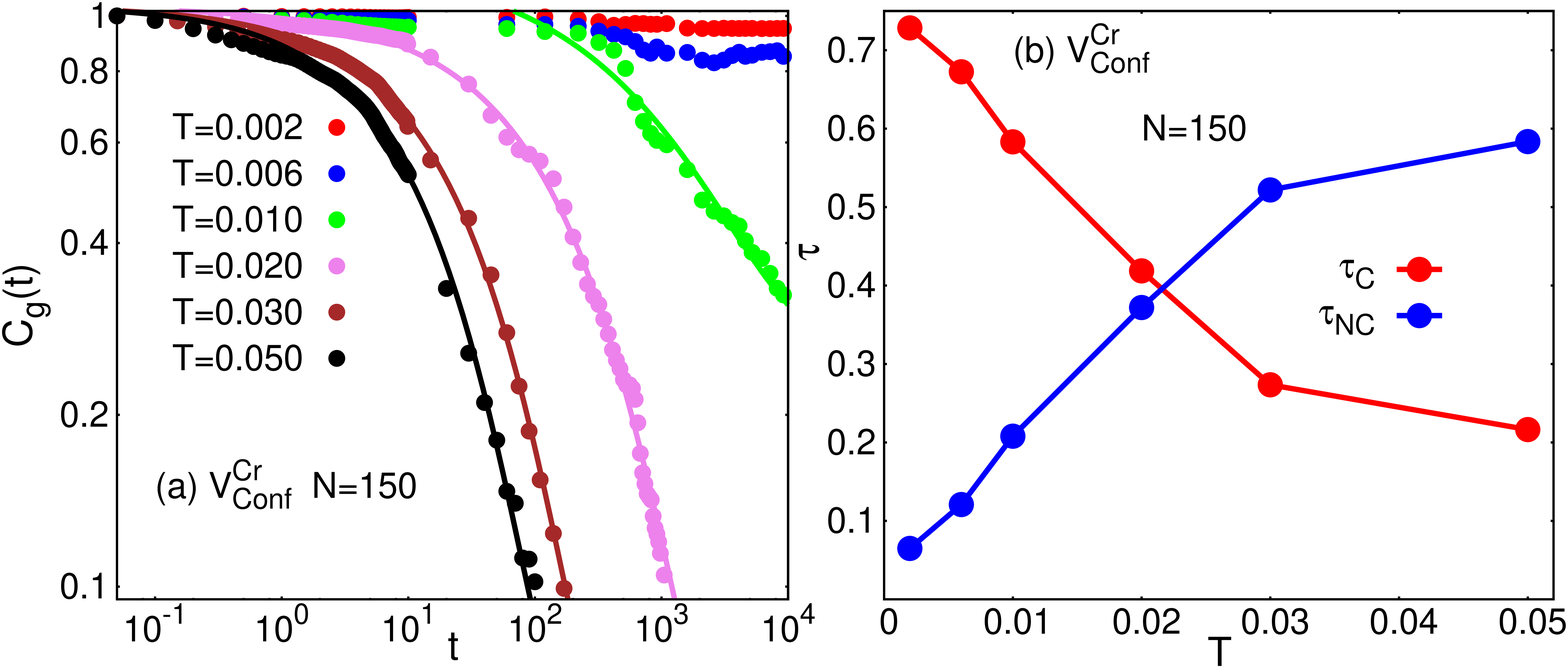}
\caption{
	 (a) Decay of cage correlation function, $c_g(t)$,  with $t$ at different $T$ for $N=150$ particles in circular confinement. 
         (b) $T-$dependence of average caging $(\tau_{C})$ and non-caging time $(\tau_{NC})$ for the same system. These two time scales cross each other at $T \approx 0.020$.
 }
\label{fig:CWM_Cg_Cng}
\end{figure}

In Fig.~\ref{fig:CWM_Cg_Cng}(a) we show the decay of cage correlation function with time at different $T$ for circular confinement. While at low $T$, $C_g(t)$ is close to unity, it shows stretched exponential decay at higher $T$. The details of the characteristic time associated with this decay is given in Table.~\ref{Table:CagrCorr}. Fig.~\ref{fig:CWM_Cg_Cng}(b) shows the $T$-dependence of average caging $(\tau_{C})$ and non-caging time $(\tau_{NC})$ for circular confinement. For $V_{\rm conf}^{\rm Cr}$, we find that the fraction of solid like particles is quite higher compared to $V_{\rm conf}^{\rm Ir}$ at low $T$ (see Fig. 8(b)). We also find that these two time scales cross each other around the crossover temperature, $T_X \approx 0.020$, as discussed in main paper.

\section{Distribution of persistence and exchange times}

\begin{figure}[ht!]
\includegraphics[width=10cm,keepaspectratio]{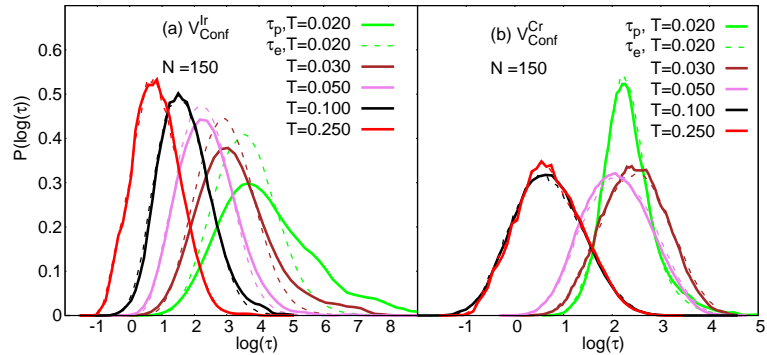}
\caption{
	 Distribution of persistence and exchange times for $N=150$ particles in (a) irregular and (b) circular confinements. The decoupling of the two distributions at low $T$ demonstrates the qualitative similarity of the particle dynamics in irregular traps with those in glassy systems. Such decoupling is not observed for circular traps.
 }
\label{fig:CWM_Pe_Ex}
\end{figure}

Fig.~\ref{fig:CWM_Pe_Ex} shows the distribution of persistence and exchange times for irregular (panel a) and circular (panel b) confinements with $N=150$ particles. For $T>0.030$, the exchange $(\tau_e)$ and persistence $(\tau_p)$ time distributions coincide for both the confinements. As $T$ becomes close to $T_X (\approx 0.020)$, the two distributions become distinct for irregular confinement; mean value for the distribution of persistence time moves toward longer times (Fig.~\ref{fig:CWM_Pe_Ex}(a)). Thus, we see the signature of decoupling, similar to what is described in main text for $N=500$, even for smaller systems. For circular confinement, while there is no sign of decoupling between $\tau_e$ and $\tau_p$, but for $T \ge 0.10$, all the distributions coincide as shown in Fig.~\ref{fig:CWM_Pe_Ex}(b).

Thus, from our analysis we find that, the static properties of Coulomb interacting particles in circular and irregular confinements show qualitatively similar temperature dependence but the dynamical responses make clear distinction between the two. Signatures of slow and heterogeneous dynamics, while present in two systems, the qualitative measures differ substantially. Azimuthal symmetry of the $V_{\rm conf}^{\rm Cr}$ leads to circularly symmetric spatially heterogeneous dynamics of the particles at low temperatures. 

\end{document}